\documentclass[english, usenatbib]{mn2e}
\usepackage{epsfig}
\usepackage{fix2col}
\usepackage{lmodern}
\usepackage[T1]{fontenc}
\usepackage[latin9]{inputenc}
\usepackage[a4paper]{geometry}
\usepackage{units}
\usepackage{soul}
\usepackage{amsmath}
\usepackage{amssymb}


\usepackage{babel}

\title[Quenching and Morphological Transformation]
      {Quenching and Morphological Transformation in Semi-Analytic Models and CANDELS}
      \author[Brennan et al.]{Ryan Brennan$^{1}$\thanks{E-mail:
          brennan@physics.rutgers.edu}, Viraj Pandya$^{2}$, Rachel
        S. Somerville$^{1}$, Guillermo Barro$^{3}$, \and Edward N. Taylor$^{4}$, Stijn Wuyts$^{5}$, Eric F. Bell$^{6}$, Avishai Dekel$^{7}$, \and Henry C. Ferguson$^{8}$, Daniel H. McIntosh$^{9}$, Casey Papovich$^{10}$, Joel Primack$^{11}$\\ $^{1}$Department of
        Physics and Astronomy, Rutgers, The State University of New
        Jersey, 136 Frelinghuysen Rd, Piscataway, NJ
        \\ $^{2}$Department of Astrophysical Sciences, Peyton Hall,
        Princeton University, Princeton, NJ \\ 
$^{3}$UCO/Lick Observatory, Department of Astronomy
        and Astrophysics, University of California, Santa Cruz, CA \\ 
$^{4}$School of Physics, the University of Melbourne,
        Parkville, VIC 3010, Australia \\ $^{5}$Max-Planck-Institut f{\"u}r extraterrestrische
        Physik, Giessenbachstrasse 1, D85748 Garching bei M{\"u}nchen,
        Germany \\ 
$^{6}$Department of Astronomy, University of Michigan, Ann Arbor, MI, USA \\ 
$^{7}$Center for Astrophysical and Planetary Science, Racah Institute of Physics, The Hebrew University, Jerusalem 91904, Israel \\ 
$^{8}$Space Telescope Science Institute, 3700 San Martin Drive, Baltimore, MD 21218, USA \\ $^{9}$Department of Physics and Astronomy, University of Missouri-Kansas City, 5110 Rockhill Road, Kansas City, MO 64110, USA \\ $^{10}$George P. and Cynthia Woods Mitchell Institute for Fundamental Physics and Astronomy, and Department of Physics and Astronomy, \\\hspace{3.5 mm}Texas A\&M University, College Station, TX 77843-4242, USA \\ $^{11}$Department of Physics, University of California at Santa Cruz, Santa Cruz, CA 95064, USA}

\begin{document}
\date{}

\maketitle

\label{firstpage}
\begin{abstract}
We examine the spheroid growth and star formation quenching
experienced by galaxies since $z\sim3$ by studying the evolution with
redshift of the quiescent and spheroid-dominated fractions of galaxies
from the CANDELS and GAMA surveys. We compare the observed fractions
with predictions from a semi-analytic model which includes
prescriptions for bulge growth and AGN feedback due to mergers and
disk instabilities. We facilitate direct morphological comparison by
converting our model bulge-to-total stellar mass ratios to S{\'e}rsic
indices. We then subdivide our population into the four quadrants of
the sSFR-S{\'e}rsic index plane and study the buildup of each of these
subpopulations. We find that the fraction of star forming disks
declines steadily, while the fraction of quiescent spheroids builds up
over cosmic time. The fractions of star forming spheroids and
quiescent disks are both non-negligible, and stay nearly constant over
the period we have studied. Our model is qualitatively successful at
reproducing the evolution of the two ``main'' populations (star
forming disks and quiescent spheroids), and approximately reproduces
the relative fractions of all four types, but predicts a stronger
decline in star forming spheroids, and increase in quiescent disks,
than is seen in the observations. A model with an additional channel
for bulge growth via disk instabilities agrees better overall with the
observations than a model in which bulges can grow only through
mergers. We also examine the relative importance of these different
physical drivers of transformation (major and minor mergers and disk
instabilities).
\end{abstract}


\section{Introduction}

The mechanisms by which galaxies are transformed and evolve over time,
both in terms of their star formation rates and their morphologies,
are still not clearly known. At low redshift, the distribution of
galaxy colors is bimodal \citep{Baldry2004,Bell2004b}. This division
of galaxies into the star forming ``blue cloud'' and the quiescent
``red sequence'' can be observed most clearly in the color-magnitude
and specific star formation rate (sSFR)-stellar mass planes
\citep{Baldry2004,Brinchmann2004,Kauffmann2003,Strateva2001}.
Additionally, star forming galaxies can be said to occupy a ``star
forming main sequence,'' a correlation between the star formation rate
and the stellar mass of star forming galaxies
\citep{Noeske2007,Daddi2007,Elbaz2007,Rodighiero2011}.  Galaxies that
are part of the red sequence have a wider range of star formation
rates, although they do exhibit a correlation between mass (or
luminosity) and color, where more massive galaxies tend to be redder
\citep{Bernardi2003,Gallazzi2006, Peng2010,Brammer2011,
  Muzzin2013}. In addition to the bimodality due to stellar
populations, there is also a bimodality in the structure of galaxies
\citep{Kauffmann2003}.  Often characterized by the bulge-to-total
luminosity or mass ratio, or light profile parameterizations such as
the S{\'e}rsic index, galaxy morphology tends to be correlated with
the star formation activity in the galaxy. Galaxy disks tend to be
bluer than bulges \citep{Peletier1996, Bell2004} and galaxies that are
part of the blue cloud are more likely to be disk-dominated, while
galaxies that are members of the red sequence are more likely to have
more prominent bulges, or to have the concentrated light profiles that
are characteristic of early type galaxies
\citep{Blanton:2009,Schiminovich2007,Bell2008,Cheung2012}.

Large surveys have shed light on how the galaxy population evolved
over a large fraction of the age of the universe. These observations
have shown that the bimodality seen in the local universe is in
place even at $z\sim2-3$ \citep{Brammer2009,Brammer2011,Muzzin2013}.
Analysis of the buildup of stellar populations from high redshift to
the present reveals that the stellar mass contained in objects in the
blue cloud has remained relatively constant, while the stellar mass
represented by galaxies on the red sequence has grown significantly;
this implies that blue star forming galaxies are in fact being
transformed into red, quiescent ones \citep{Bell2004b, Borch2006,
  Bell2007,Faber2007}. The mechanism responsible for this
``quenching'' (or turning off of star formation) is not so clear. New
information about the evolution of galaxy structure and morphologies
has recently been gleaned from observations using the Hubble Space
Telescope. Recent work suggests that quiescence is intimately tied
to the presence of a bulge component \citep{Wuyts2011,Lang2014,Bluck2014,McIntosh2014}. Moreover, observations have revealed
a population of compact spheroid-dominated star forming galaxies at
$z\sim 2$, which may be the progenitors of the quiescent, elliptical
galaxies we see today \citep{Wuyts2011,Whitaker2012,Barro2013,
  Williams2014, Barro2014, Williams2014b}.  It seems likely that the
mechanisms responsible for quenching, morphological, and size
evolution are connected.

There have been several mechanisms proposed to explain galaxy
quenching.  One of the most popular scenarios involves feedback due to
active galactic nuclei (AGN). AGN feedback can be broadly divided into
two regimes: the radiatively efficient ``quasar'' or ``bright'' mode,
which is proposed to drive a powerful wind which expels gas from the
galaxy, and the ``radio'' or ``maintenance'' mode, which heats gas in
the galactic halo, preventing it from cooling and forming stars
\citep[][and references therein]{Somerville_Dave:2014}.  This AGN
activity can be driven either by galaxy mergers
\citep{Ellison2011,Silverman2011} or in situ processes such as disk
instabilities \citep{Bournaud2011,Dekel2014}. Both of these processes
lead to rapid transfer of angular momentum and the growth of a bulge
component.  Virial shock heating is another proposed mechanism:
during collapse, gas can be heated via the conversion of gravitational
potential energy into kinetic energy \citep{White:1978}. Above a
(redshift dependent) critical halo mass of $\sim10^{12}M_{\odot}$,
this shock heating may be able to keep a substantial fraction of the
halo gas hot, leading to quenching \citep{Birnboim2003,Keres2005}.

While this does not seem directly related to the presence of a bulge
component, it is clear from observations that galaxies residing in
halos above $10^{12}M_{\odot}$ are more likely to be bulge-dominated
than disk-dominated \citep{Dekel2009,Woo2014}.  There is also the
possibility that the presence of a (significant) bulge may itself
stabilize the disk against local instabilities, thus making star
formation less efficient, an effect known as morphological quenching
\citep{Martig2009}. Finally, there is a suite of processes connected
with dense environments, including tidal and ram pressure stripping
and harrassment. These are often collectively referred to as
``environmental quenching'' \citep{Oemler1974, Dressler1980,
  Balogh2004b, Tinker2010,Peng2010}, and they likely primarily affect
satellites orbiting within a larger halo. These processes probably
operate on a different timescale, and lead to different sorts of
morphological transformation, than the ones described above. In this
paper, we focus on field galaxy environments, so environmental
processes are likely to be sub-dominant. See also \citet{McIntosh2014}
for a summary of proposed quenching processes.

If quenching and morphological transformation are (in most cases)
intimately tied to each other, then galaxies which seem to be the
``outliers'' in this picture may be of particular interest: the
quiescent disk-dominated galaxies and star forming spheroid-dominated
galaxies. These populations are smaller than those of star forming
disk-dominated and quiescent spheroid-dominated galaxies, although
they are not insignificant in size
\citep{McGrath2008,vanderWel2011}. \citet{Schawinski2014} did an
analysis of galaxies in the local universe that occupy the ``green
valley,'' the region in between the blue cloud and the red sequence on
the color-magnitude diagram, using observational data from SDSS
\citep{York2000} and GALEX \citep{Martin2005}. They used morphology
classifications from Galaxy Zoo \citep{Lintott2008,Lintott2011} and
determined that there were two distinct paths through the green
valley, one taken by galaxies that leave the blue cloud as disk
dominated systems, the other by galaxies that transition as
bulge-dominated systems (see also the work of
  \citep{Smethurst2015}. The path associated with spheroid-dominated
galaxies is consistent with work that suggests that bulge growth
precedes quiescence \citep{Wuyts2011,Wong2012, Lang2014}, while the path taken
by disk-dominated galaxies may explain the slowly growing population
of quiescent disk galaxies in the local universe which have been cut
off from their gas supply but suffered nothing catastrophic to destroy
or use up their existing gas reservoirs. \citet{Barro2013} and \citet{Woo2014} similarly
identify different scenarios of quenching based on structural
evolution. \citet{McIntosh2014} also describe a path associated with
spheroid-dominated galaxies when examining what they call ``recently
quenched ellipticals'' in the Sloan Digital Sky Survey. In this light,
these ``outlying'' populations are possibly much more important to the
overall picture than originally suspected, and may be an indication
that quenching is caused by multiple physical processes.

Aiding in the investigation of galaxy formation and evolution are
state of the art numerical simulations and semi-analytic
models. N-body dark matter simulations such as Bolshoi
\citep{Klypin2011} are an invaluable tool when testing the predictions
of our currently favored Lambda Cold Dark Matter cosmological model,
$\Lambda$CDM. Semi-analytic models, or SAMs, plant galaxies in merger
trees assembled in dark matter N-body simulations or constructed using
techniques based on the Extended Press Schechter formalism. By
following the evolution of these galaxies within the backbone of the
dark matter history, accounting for physical processes such as gas
accretion and cooling, star formation, merging and feedback with
physically motivated recipes, population statistics for a cosmological
sample of galaxies can be generated quickly and with minimal
computational resources
\citep{Kauffmann1993b,Cole2004,Somerville1999,DeLucia2006,Somerville2008,Guo2011,Somerville2012,
  Porter2014}.  SAMs have been used to study the evolution of star
formation and the buildup of spheroid-dominated galaxies and to gauge
which processes are especially important. \citet{DeLucia2006} and
\citet{Benson2010} have investigated the buildup of spheroid-dominated
galaxies in SAMs with cosmic time. These analyses and others have
reiterated that bulge growth often appears to be connected to the
cessation of star formation and also demonstrate that the two main
channels for bulge growth are mergers and disk instabilities, both of
which appear to be important, although their degree of importance may
change with redshift, galaxy mass and environment, and is also model
dependent \citep{Parry2009,DeLucia2011,Fontanot2012,Porter2014}.

Recently, \citet{Porter2014} compared the predictions of the latest
version of the ``Santa Cruz'' SAM
\citep{Somerville1999,Somerville2008} for the $z=0$ stellar mass
function divided by morphology with available observations, and found
fairly good agreement. They found that adding a prescription for bulge
growth via disk instabilities brought the model into better agreement
with the observed galaxy stellar mass function of spheroid-dominated
galaxies at intermediate masses
($10.5<\rm{log}(M_{*}/M_{\odot})<11.5$). In addition,
\citet{Porter2014} developed a new model for predicting the radial
\emph{sizes} and velocity dispersions of bulges formed via mergers or
disk instabilities, based on a simple analytic model calibrated using
numerical hydrodynamic simulations of binary galaxy mergers. Their
model reproduces the observed size-mass relation for spheroids and
disks, and the evolution of this relation from $z\sim 2$ to the
present day (see also Somerville et al. in prep). \citet{Porter2014b}
investigated the predictions of the same models for the correlation of the
age and metallicity of stars in local spheroid-dominated galaxies with
structural parameters such as size and velocity dispersion. They found
a strong correlation between both stellar population parameters (age
and metallicity) and internal velocity dispersion, in agreement with
observations. They found no correlation between age and radius, and a
weak correlation between metallicity and radius, also in agreement
with observations of nearby early type galaxies. In this paper we
follow up on the work by Porter and collaborators by directly studying
the build-up of the spheroid-dominated population over cosmic time,
and comparing with observations of high-redshift galaxies.

The conclusions of previous studies in the literature regarding the
spheroid-dominated fraction of galaxies in SAMs and the agreement with
observations are difficult to synthesize, because different analyses
use different criteria to define spheroid-dominated galaxies both in
the models and in the observations. The bulge-to-total mass ratio
($B/T$) that is readily predicted in SAMs is difficult to measure
observationally. Therefore it has been difficult to make a direct
comparison between model predictions and observations previously. One
of the important new features of this study is that we extend our
models to predict a morphological quantifier that can be compared more
directly with observations. We describe our new method in detail
below.

In this paper, we present new results quantifying the evolution of
quenching and spheroid growth in observations from $z\sim3$ to the
present, and also present new predictions of these same quantities
from state-of-the-art semi-analytic models. We split galaxies
according to their star formation rates and morphologies and examine
the buildup of the quiescent and spheroid-dominated fractions of
galaxies. We then go further than studies in the past by subdividing
into four populations: star forming disk-dominated galaxies (SFDs),
star forming spheroid-dominated galaxies (SFSs), quiescent
disk-dominated galaxies (QDs) and quiescent spheroid-dominated
galaxies (QSs). We examine the evolution of the fraction of galaxies
in each of these populations.  Our low redshift observational data ($z
\sim 0.06$) come from the Galaxy and Mass Assembly Survey
\citep[GAMA][]{Driver2009} and our higher redshift data ($0.5<z<3.0$)
come from the Cosmic Assembly Near-infrared Deep Extragalactic Legacy
Survey (CANDELS; \citealt{Grogin2011, Koekemoer2011}). We use the Santa
Cruz SAM of \citet{Somerville2008} with updates as described in
\citet{Somerville2012} and \citet{Porter2014}. Our semi-analytic model
includes the effects of AGN feedback and bulge growth triggered by
mergers and (optional) disk instabilities.  Another way in which our
study is unique is the way we characterize the morphologies of our
model galaxies: we convert our model output, bulge-to-total mass
ratio, to S{\'e}rsic index as described in Section 2 and Appendix A in
order to facilitate a more direct comparison between model and
observed galaxy morphologies than has been carried out before. We will
also examine in detail the histories of galaxies selected from each
population in order to shed light on the individual tracks that
different types of galaxies move along as they evolve. The structure
of the paper is as follows. In Section 2 we give an overview of our
semi-analytic model and of the data sets with which we are comparing.
In Section 3 we present a comparison of the evolution of these
populations in the model and the observations. We present our
discussion, in part informed by studying individual evolutionary
tracks of galaxies from the model, in Section 4 and our summary and
conclusions in Section 5.

\section{Semi-Analytic Model and Observational Data}

\subsection{\emph{The Semi-Analytic Model}}

The SAMs used in this paper were first presented in
\citet{Somerville1999} and \citet{Somerville2001}, and significantly
updated in \citet[][S08]{Somerville2008},
\citet[][S12]{Somerville2012} and \citet[][P14]{Porter2014}. The model
includes prescriptions for the following physical processes: the
hierarchical growth of structure in the form of dark matter merger
trees, the heating and cooling of gas, star formation as governed by
the empirical Kennicut-Schmidt law, the evolution of stellar
populations, supernova feedback, chemical evolution of the ISM and ICM
due to supernovae, AGN feedback, and starbursts and morphological
transformation due to galaxy mergers and disk instabilities. Here we
will briefly summarize these processes, focusing mainly on the
processes relevant to the evolution of star formation and
morphology. For a more in depth description of the model, see S08 and
P14.  We assume a standard $\Lambda$CDM cosmology ($\Omega_{m}=0.27,$
$\Omega_{\Lambda}=0.73,$ $h=0.7$ ) and a \citet{Chabrier2003} initial
mass function. Our adopted baryon fraction is 0.1658. Our cosmology
was chosen to match that adopted by the Bolshoi simulation (detailed
below) and is consistent with the Wilkinson Microwave Anisotropy Probe
(WMAP) 5/7-year results \citep{Komatsu2009,Komatsu2011}.

In this work, we use the CANDELS lightcones (Somerville et al. in
prep) extracted from the Bolshoi dark-matter only N-body simulation
\citep{Klypin2011,Trujillo2011}. Dark matter halos are identified
using the ROCKSTAR algorithm of \citet{Behroozi2013}. The Bolshoi
simulation is complete down to halos with $V_{\rm{circ}}=50$ km/s, and
has a force resolution and mass resolution of $1$$h^{-1}$ kpc and
$1.9\times10^{8}$$M_{\odot}$, respectively. Merger trees are
constructed for each halo in the lightcone using the method of
\citet{SK:1999}. There is no appreciable difference in results when
using merger trees extracted from the N-body simulation (as done in
Porter et al. 2014) as opposed to EPS (as we do here). For our lowest redshift bin,
the lightcones represent a very small volume so we simply use a low-z
snapshot from the Bolshoi volume.

When dark matter haloes merge, the central galaxy of the largest
progenitor becomes the new central galaxy, while all other galaxies
become satellites.  Satellite galaxies are able to spiral in and merge
with the central galaxy, losing angular momentum to dynamical friction
as they orbit.  The merger time-scale is estimated using a variant of
the Chandrasekhar formula from \citet{Boylan2008}. Tidal stripping and
destruction of satellites as described in S08 are also included.

Before the universe is reionized, each halo has a hot gas mass equal
to the virial mass of the halo times the universal baryon fraction.
The collapse of gas into low-mass haloes is suppressed after
reionization due to the photoionizing background. We assume the
universe is fully reionized by $z=11$ and use the results of
\citet{Gnedin2000} and \citet{Kravtsov2004} to model the fraction of
baryons that can collapse into haloes of a given mass following
reionization. Due to the galaxy mass range selected in this work, we
do not expect our results to be sensitive to this prescription.

When dark matter haloes collapse or are involved in a merger that at
least doubles the mass of the progenitors, the hot gas is shock-heated
to the virial temperature of the new halo. The rate at which this gas
can cool is determined by a simple spherical cooling flow
model. Assuming a monotonically decreasing density profile for the
gas, and that denser gas cools faster, we can define a ``cooling
radius'', within which all gas is able to cool within some time
$t_{\rm{cool}}$, which we have defined as the halo dynamical time. The
initial density profile is assumed to be that of a singular isothermal
sphere, and the cooling radius is found by using the atomic cooling
curves of \citet{Sutherland1993}.  The cooling radius may be larger or
smaller than the virial radius of the halo; when the cooling radius is
larger, the cooling rate is limited only by the rate at which gas is
infalling. The transition from $r_{\rm{cool}}>r_{\rm{vir}}$ to
$r_{\rm{cool}}<r_{\rm{vir}}$ is associated with the transition from
``cold flows'', where cold gas streams into the halo along dense
filaments without being heated, to ``hot flows'', where gas is shock
heated on its way in, forming a diffuse hot gas halo before cooling
\citep{Birnboim2003,Dekel2006,Keres2005}. Note that in this way,
virial shock heating (sometimes referred to as `halo mass quenching')
is included in our SAMs. However, it has been shown by many studies
(both numerical and semi-analytic) that this effect alone is
insufficient to create the observed population of massive quiescent
galaxies \citep[][and references therein]{Somerville_Dave:2014}.

Newly cooled gas collapses to form a rotationally supported disk, the
scale radius of which is estimated based on the initial angular
momentum of the gas and the profile of the halo. We assume that
angular momentum is conserved and that the self-gravity of the
collapsing baryons causes the inner part of the halo to contract
\citep{Blumenthal1986,Flores1993,Mo1998}.  This method was shown to
reproduce the observed size-stellar mass relations of disks out to
$z\sim2$ in \citet{Somerville2008b}. Spheroids can be created by
mergers or disk instabilities. The sizes of spheroids formed in
mergers are determined by the stellar masses, sizes and gas fractions
of the two progenitors, as described in P14. The size of spheroids
formed in disk instabilities is determined by assuming that they form
from the center of the exponential stellar disk; the radius is simply
the radius that contains the amount of mass that is to be transferred
from the disk to the bulge (again, see P14 for details).

There are two modes of star formation in the model: a ``normal'' mode
that occurs in isolated disks and a ``starburst'' mode that occurs as
a result of a merger or internal disk instability, which will be
discussed in more depth below. The normal mode follows the
Schmidt-Kennicutt relation \citep{Kennicutt1998} and assumes that gas
must be above some fixed critical surface density (the adopted value
here is $6\, M_{\odot}/\rm{pc}^{2}$) in order to form stars.

Exploding supernovae and massive stars are capable of depositing
energy into the ISM, which can drive outflows of cold gas from the
galaxy.  We assume that the mass outflow rate is proportional to the
SFR and decreases with increasing galaxy circular velocity, in
accordance with the theory of ``energy-driven'' winds. Some ejected
gas is removed from the halo completely, while some is deposited into
the hot gas reservoir of the halo and is eligible to cool again. The
gas that is driven from the halo entirely is combined with the gas
that has been prevented from cooling by the photoionizing background
and may later reaccrete back into the halo. The fraction of gas which
is retained by the halo versus the amount that is ejected is a
function of halo circular velocity as decribed in S08.

Heavy elements are produced by each generation of stars, and chemical
enrichment is modelled simply using the instantaneous recycling
approximation. For each parcel of new stars $dm_{*}$, a mass of metals
$dM_{\rm{Z}}=ydm_{*}$ is also created, which is immediately mixed with
the cold gas in the disk. The yield $y$ is assumed to be constant
and is treated as a free parameter. Supernova driven winds act to
remove some of this enriched gas, depositing a portion of the created
metals into the hot gas or outside of the halo.

\subsubsection{Mergers and Starbursts}

Mergers between galaxies are assumed to remove angular momentum from
stars and gas in the disk and drive material towards the center,
building up a spheroidal component. In our model, this spheroidal
component is formed instantaneously. In principle this could affect
our results by forming bulges more quickly than they should form in
the real universe. However, actual bulge formation time scales
($\sim$$t_{\rm{dyn}}$) are quite short compared with the times
associated with our redshift bins, so we don't expect this to have
much of an effect.

Mergers also trigger a starburst, the efficiency of which depends on
the gas fraction of the central galaxy and the mass ratio of the two
progenitors. The time scale of the burst is also determined by
properties of the progenitor galaxies. The parameterization is based
on hydrodynamical simulations of binary mergers between disks
\citep{Hopkins2009a}. Simulations show that the closer the mass ratio
of the progenitors is to one (or how ``major'' the merger is) and the
more gas-poor the merger is, the more efficient it is at removing
angular momentum from the gas and driving it into the nucleus, and
scattering disk stars into a hot spheroid component
\citep{Cox2006,Robertson2006}. The gas fraction dependence can be
understood as follows: if the progenitors are very gas-rich, there is
not enough stellar mass to create a torque on the gas, making it
difficult for the gas to shed angular momentum and collapse inward
\citep{Hopkins2009a}. S08 parameterized the burst efficiency only as a
function of mass ratio, but S12 and P14 introduced the gas fraction
dependence in accordance with \citet{Hopkins2009b}. Stars that are
formed as part of the starburst are added to the spheroidal component,
as are 80\% of the stars from the merging satellite galaxy.  The other
20\% are deposited into a diffuse stellar halo component.

\subsubsection{Disk Instabilities}

Disk material can also be converted into a spheroidal component as a
result of internal gravitational instabilities.  A pure disk without a
dark matter halo is very unstable to the formation of a bar or bulge,
while massive dark matter haloes tend to stabilize a thin, cold
galactic disk \citep{Ostriker1973,Fall1980}. When the ratio of dark
matter mass to disk mass falls below a critical value, the disk can no
longer support itself and material collapses into the inner regions of
the galaxy \citep{Efstathiou1982}. Here we adopt an avenue for bulge
growth due to disk instability, based on a Toomre-like stability
criterion.  Following \citet{Efstathiou1982}, \citet{Mo1998}, P14 and
many other works, we define the stability parameter as

\begin{eqnarray}
\epsilon_{\rm{disk}}=\frac{V_{\rm{max}}}{(GM_{\rm{disk}}/r_{\rm{disk}})^{\nicefrac{1}{2}}}
\end{eqnarray}

where $V_{\rm{max}}$ is the maximum circular velocity of the halo
(used as a proxy for the maximum circular velocity of the disk),
$r_{\rm{disk}}$ is the scale length of the stellar disk and
$M_{\rm{disk}}$ is the stellar mass of the disk. This is identical to
the ``\textbf{Stars DI}'' disk instability criterion introduced in
P14. Whenever $\epsilon_{\rm{disk}}<\epsilon_{\rm{crit}}$, the disk is
considered to be unstable.  The value of $\epsilon_{\rm{crit}}$ in
numerical simulations of isolated disks has been found to be in the
range of $0.6-1.1$, with disks containing stars and cold gas having a
lower threshold than pure stellar disks \citep{Efstathiou1982,Mo1998}.
We set $\epsilon_{\rm{crit}}=0.75$ as in P14, where this value was
chosen to match the observed fraction of spheroid-dominated galaxies
at $z=0$.  When the disk becomes unstable, stellar mass is moved from
the disk to the bulge until
$\epsilon_{\rm{disk}}=\epsilon_{\rm{crit}}$.  The gas in the disk is
not affected. The ``\textbf{Stars+Gas DI}'' model of P14 included gas
in determining the stability of the disk and also moved some gas to
the bulge component to feed the central supermassive black hole when
the disk became unstable. However, the results for the two approaches
were very similar. Again, the creation of the bulge component is
instantaneous. While we are aware that this implementation of disk
instability is crude and perhaps does not capture all of the relevant
physics, this is an approach that is commonly used in the
literature. One of the goals of this work is to explore how important
bulge growth through disk instabilities might be, in order to guide
future investigations. Later, we discuss more physical models of disk
instability and how including them might affect the results of this
study.

It is worth noting that we also do not account for the possibility
that a previously existing bulge may help stabilize the disk against
another instability. Because of this (and the fact that we only move
as much material as needed to restabilize the bulge) it is possible
(even common) for disks to develop chronic instabilities which lead to
the steady growth of a bulge component.

Below we present our results for versions of the SAM both with the
disk instability prescription turned on (DI model) and off (noDI
model). The DI model is our fiducial model, however, and unless
otherwise noted, it is the DI model that is shown.

\subsubsection{Black Hole Accretion and Feedback}

Galaxies are initially seeded with a massive black hole of $10^{4} \,
M_{\odot}$ \citep{Hirschmann2012}.  When two galaxies merge as
described above, their central black holes are assumed to merge as
well, after which the new central black hole of the merger remnant
engages in a bout of feeding and radiatively efficient, or ``quasar''
mode, AGN activity. During this time, the black hole accretes at its
Eddington limit. As the black hole accretes and radiates, it deposits
energy into the surrounding medium until it reaches a critical mass
which corresponds to the energy which would stop accretion and begin
driving an outflow, such as those seen in many recently merged
  systems\citep{Rupke2013, Emonts2014}. The black hole effectively
starves itself of material, as its accretion rate declines as a power
law, in accordance with the results of \citet{Hopkins2006}. We follow
the hydrodynamical binary merger simulations of \citet{Hopkins2007}
for our definition of the critical mass, $M_{\rm{crit}}$, at which the
black hole accretion rate enters the declining phase, and
$M_{\rm{final}}$, at which the black hole stops feeding. If the newly
merged black hole is already more massive than $M_{\rm{final}}$, there
is no accretion event. We note that our predicted final black hole and
bulge masses are consistent with the observed
$M_{\rm{BH}}-M_{\rm{bulge}}$ relation
\citep{Somerville2008,Hirschmann2012}.

A bout of black hole accretion and AGN activity can also be triggered
by a disk instability. When disk mass is transferred to the bulge as
previously described, we assume the black hole accretes a gas mass
equivalent to some fraction of that mass. Following
\citet{Hirschmann2012}, we set this term to be
$f_{\rm{fuel,DI}}=0.002$, which leads to good agreement with the
observed number density of low-luminosity AGN. The black hole can
continue to accrete until this fuel is consumed.

The black hole is also able to feed and effect feedback in the
``radio'' or ``maintenance'' mode. In this mode the black hole feeds
via Bondi-Hoyle accretion from the hot halo \citep{Bondi1952}. The
accretion is usually significantly sub-Eddington. This feedback mode
is associated with giant radio jets which heat the surrounding gas,
preventing it from cooling and forming stars. Once the accretion rate
is determined, a coupling constant determines how effectively the
energy released couples to the surrounding gas. The radio mode heating
rate is then calculated and subtracted from the cooling rate described
above.

\subsubsection{Computing S{\'e}rsic Indices}

This work involves comparing the morphologies of model galaxies with
$M_{*}>10^{10}M_{\odot}$ with those of observed galaxies.  From the
SAM, we can easily calculate the bulge-to-total stellar mass ratio,
and we have predictions for the radii of the galactic bulge and
disk. There are many different methods used to classify the
morphologies of observed galaxies. One commonly used method is to fit
the light profile with a S{\'e}rsic function, resulting in the
determination of the S{\'e}rsic index \citep{Sersic1963}.  In an
effort to put the observations and model on equal footing, we have
converted our model outputs to a S{\'e}rsic index using a lookup table
which takes in the bulge-to-total mass ratio and bulge radius to disk
radius ratio and gives an effective radius and S{\'e}rsic index. This
lookup table was generated by fitting S{\'e}rsic indices and effective
radii to synthetic bulge+disk systems ($n$=1 for disks and $n$=4 for
bulges) for a range of different bulge-to-total mass ratios and bulge
radius to disk radius size ratios. The values that come out of the
lookup table are discrete for obvious reasons, so we use a 2D
interpolation of the table to generate our S{\'e}rsic indices and
effective radii. More information and some tests of our approach can
be found in Appendix A.

\subsection{\emph{Observational Samples}}
\subsubsection{CANDELS}

In this work, we make use of \emph{Hubble Space Telescope/Wide Field
  Camera 3} (HST/WFC3) observations of galaxies taken as part of the
Cosmic Assembly Near-infrared Deep Extragalactic Legacy Survey
(CANDELS; \citealt{Grogin2011, Koekemoer2011}). These observations
span two of the five CANDELS fields: the Ultra-Deep Survey (UDS;
\citealt{Lawrence2007}) and the Great Observatories Origins Deep Survey
South (GOODS-S; \citealt{Giavalisco2004}). With the combined strengths
of galaxy selection in the F160W (H) band and the availability of rich
multi-wavelength datasets, the CANDELS catalogs afford us an
unprecedented study of galactic structure and the rise and fall of
star formation activity from high redshift toward the present day.

Here, we give only an outline of the galaxy catalogs used for our analysis\footnote{All CANDELS catalogs are available at the Rainbow Database-http://arcoiris.ucolick.org/Rainbow\_navigator\_public/}. The galaxies were drawn from the CANDELS catalogs for UDS and GOODS-S; for more details we refer the reader to
\citet{Galametz2013} and \citet{Guo2013}, respectively. Briefly, the UV-to-NIR multiwavelength photometric catalogs were computed using the
template-fitting method TFIT \citep{Lee2012, Laidler2007} which allows us to consistently merge datasets with significantly different spatial resolution. 

Photometric redshifts were determined following the method described in \citet{Dahlen2013} which combines redshift probability distributions from several different codes using a Bayesian approach to improve the precision and reduce the number of catastrophic outliers. For a sample of 480
spectroscopically confirmed GOODS-S galaxies without an active
galactic nucleus, the median and standard deviation of
$(z_{\textrm{phot}}-z_{\textrm{spec}})/(1+z_{\textrm{spec}})$ are
approximately -0.014 and 0.045, respectively.

The stellar masses were drawn from the catalog presented in \citet{Santini2014}. The catalog includes stellar masses computed using different SED fitting codes and modeling assumptions. For this work, we adopted stellar masses computed using FAST \citep{Kriek2009} with \citet{BC2003} stellar population synthesis models, a Chabrier initial mass function \citet{Chabrier2003}, exponentially declining star formation histories, solar metallicity and the Calzetti dust extinction law \citep{Calzetti2001}. The total star formation rate for galaxies
detected in the mid- and/or far-infrared is defined as
SFR$_{\mathrm{total}}$=SFR$_{\mathrm{UV+IR}}$ =
SFR$_{\mathrm{UV}}$+SFR$_{\mathrm{IR}}$, where
SFR$_{\mathrm{UV}}\equiv$ SFR$_{\mathrm{2800}}$ is the unobscured (and
therefore uncorrected for dust extinction) component derived from
$L_{\mathrm{2800}}$ (the luminosity at 2800\AA), and
SFR$_{\mathrm{IR}}$ is the integrated $L_{\mathrm{IR}}$-based obscured
component. The integrated (or \emph{total}) infrared luminosity,
$L_{\mathrm{IR}}$, is itself derived from fitting 24 $\mu$m fluxes
using \citet{Chary2001} templates and a calibration determined from
\emph{Herschel} data (see \citealt{Elbaz2011}). Two major assumptions
underlying this mapping from 24 $\mu$m flux to the total IR (8$\mu$m
to 1000$\mu$m) luminosity are (1) the IR SEDs of galaxies do not
evolve significantly with redshift, and (2) emission from dust heated
by an obscured AGN does not significantly increase the 24 $\mu$m
fluxes (again, consult \citet{Elbaz2011} for the validity of these
assumptions).

To be more concrete, SFR$_{\mathrm{UV+IR}} =
1.09\times10^{-10}(L_{IR}+3.3L_{2800})$, based on
\citet{Kennicutt1998} and \citet{Bell2005}, and assuming a
\citet{Chabrier2003} IMF (see \citet{Barro2011} for more information).
For galaxies without an infrared detection and thus no total
$L_{\mathrm{IR}}$ estimate, we instead corrected SFR$_{\textrm{UV}}$
for dust extinction assuming the Calzetti law \citep{Calzetti2001},
giving us a comparable estimate of SFR$_{\mathrm{total}}$:
SFR$_{\mathrm{UV,corr}} =$ SFR$_{\mathrm{UV}}\cdot 10^ {0.4 \cdot
  1.8\cdot A_V}$. The optical extinction parameter $A_V$ for each
galaxy was determined by FAST, and the factor of 1.8 in the exponent
is the attenuation parameter, $\kappa({\lambda=2800})$. We refer the
reader to \citet{Wuyts11b} and \citet{Perez2008} for further details
about the derivation of total star formation rates, including a wealth
of comparisons between different estimates of SFR$_{\mathrm{total}}$.

The structural measurements of the observed galaxies (S{\'e}rsic
indices, in particular), as computed by GALFIT
\citep{Peng2002,vanderwel2012} from the \emph{HST/WFC3} F160W (H-band)
images, are used throughout this work. To secure the robustness and
completeness of our sample, we consider only those galaxies with
$m_{\textrm{F160W}}<25$, $M_*>10^{10}M_{\odot}$, and GALFIT flag equal
to 0 (good fits only). For the redshift range considered in this paper
($0.5\leq z\leq3.0$), these three major selection cuts leave us with
the following numbers of galaxies in each field: 1123 in GOODS-S and
1594 in UDS. The GALFIT high-quality flag cut was applied last; our
sample sizes before that particular cut were 1333 and 1798 galaxies in
GOODS-S and UDS, respectively. \citet{Guo2013} and
\citet{Galametz2013} respectively show that the GOODS-S and UDS
samples are complete after taking into account our
m$_{\textrm{F160W}}$ cut. \citet{Guo2013} further show that the
completeness in GOODS-S also depends on morphology by splitting a
synthetic comparison sample into disk-dominated and spheroid-dominated
subsamples (based on the S{\'e}rsic index). Nevertheless, our
morphology-dependent subsamples should continue to remain complete
given our m$_{\textrm{F160W}}$ and mass selection criteria.

\subsubsection{GAMA}
Given the small volume probed by CANDELS at $z\lesssim0.5$, and the
need to compare our high-redshift results to those obtained from a
robust low-redshift anchor point, we incorporate multi-wavelength data
from the Galaxy And Mass Assembly (GAMA) survey into our
analysis. GAMA is a large (144 deg$^2$) spectroscopic survey that
builds on the legacy of the Sloan Digital Sky Survey (SDSS;
\citealt{York2000}) and the Two-degree Field Galaxy Redshift Survey
(2dFGRS; \citealt{Colless2001}), reaching a limiting magnitude of
$r<19.8$ mag with $\gtrsim98$\% spectroscopic completeness (see
\citet{Driver2011} for a review of the first three years of GAMA). It
is this intermediate depth and high spectroscopic completeness,
combined with photometry spanning a large wavelength range \citep[1 nm
  to 1m][]{Liske2014}), in contrast to wider and shallower
spectroscopic surveys in the past, that makes GAMA a unique survey,
and an ideal complement to the CANDELS dataset. In addition to
conducting its own spectroscopic observations with the AAOmega
spectrograph on the Anglo-Australian Telescope, GAMA has assembled
existing and is pursuing new spectroscopic and imaging data in
collaboration with several other independent surveys (see
\citet{Baldry2010}, and Tables 1 and 4 of \citet{Liske2014}).

Specifically, this work makes use of Data Release 2 (DR2) from the
GAMA survey. DR2 provides, among other things, local bulk
flow-corrected redshifts (see Section 2.3 of \citealt{Baldry2012}),
stellar masses \citep{Taylor2011}, H$\alpha$-based star formation
rates \citep{Gunawardhana2013,Hopkins2013}, observed and rest-frame
photometry \citep{Hill2011}, and GALFIT structural measurements
\citep[][we use $r$-band fits]{Kelvin2012}. These data are released
for 72,225 objects distributed over three GAMA regions: two 48 deg$^2$
fields with limiting magnitude $r<19.0$ mag, and one 48 deg$^2$ field
with limiting magnitude $r<19.4$ mag, giving a total survey volume of
144 deg$^2$. \citet{Liske2014} provide an overview of DR2 for GAMA as
well as further information about the survey's progress.

For the sake of consistency, the H$\alpha$-based SFR for each GAMA
galaxy, SFR$_{\mathrm{H\alpha}}$, was converted from a Salpeter IMF
basis to a Chabrier IMF basis\citep{Bell2005}. The
SFR$_{\mathrm{H\alpha}}$ measurements provided by GAMA should not, in
principle, deviate greatly from the SFR$_{\mathrm{UV+IR}}$ and
SFR$_{\mathrm{UV,corr}}$ prescriptions that we have adopted for
CANDELS galaxies because the H$\alpha$ luminosity
($L_{\mathrm{H\alpha}}$) estimates for GAMA galaxies have been
corrected for dust extinction and Balmer stellar absorption
\citep{Gunawardhana2013}. Furthermore, we applied aperture corrections
to the stellar mass and rest-frame magnitude estimates to account for
the fraction of mass (or flux) that falls outside of the $r$-band
\citep[SExtractor AUTO][]{Bertin1996} aperture used for
aperture-matched photometry \citep{Liske2014}.

In order to ensure robustness and consistency, we apply the following
selection cuts to the GAMA DR2 catalog: $M_*>10^{10}M_{\odot}$,
$r$-band GALFIT flag equal to 0 (good fits only), and high-quality
redshifts only (``NQ''$>3$ in the DR2 catalog). As mentioned earlier,
the limiting magnitudes are $r=19.0$ mag for two fields, and $r=19.4$
mag for the third field. We consider only galaxies with local bulk
flow-corrected redshifts between 0.005 and 0.12; the lower limit is to
prevent stellar contamination in our galaxy sample (in fact, the
lowest redshift in our sample after applying the aforementioned cuts
is $z\sim0.0059$), and the latter limit is the maximum redshift at
which we are complete for $M_*\sim10^{10}M_{\odot}$ galaxies. These
selection cuts leave us with a total of 5112 GAMA DR2 galaxies. Before
the GALFIT high-quality flag cut, but after all other cuts, our GAMA
sample size was 5977.

For each galaxy in our final GAMA sample, we derive completeness
correction weights using the $V_{\mathrm{survey}}/V_{\mathrm{max}}$
weighting technique (\citealt{Schmidt1968}, and see Section 3 of
\citealt{Taylor2014}). $V_{\mathrm{survey}}$ is the total comoving
volume contained within the 144 square degree GAMA survey (taking into
account the survey geometry, and considering only our chosen redshift
slice, $0.005<z<0.12$). $V_{\mathrm{max}}$ is the maximum comoving
volume within which a given galaxy could have been detected, again
taking into account the survey geometry (i.e., using the provided
$z_{\mathrm{max,19.0}}$ and $z_{\mathrm{max,19.4}}$ for each galaxy,
corresponding to the three 48 square degree GAMA fields with different
$r$-band selection limits). $V_{\mathrm{survey}}/V_{\mathrm{max}}$
equals 1 for galaxies for which we are complete, and it is greater
than 1 for galaxies for which we are incomplete. The median weight is
1.0, and only 187 galaxies ($\sim$3\%) have a weight greater than 1.0
(with the maximum value being about 35). As expected, our GAMA sample
is $\sim$97\% complete.

\section{Results}

We now examine how well our model (with the disk instability
prescription turned both on and off) matches the buildup of the
quiescent and spheroid-dominated fraction of observed galaxies. We
then subdivide the model and observed populations further and examine
the buildup of the four quadrants of the sSFR-S{\'e}rsic index
plane. In this way we can assess where our model is succeeding and
failing in transforming galaxies in terms of their star formation
rates and morphologies.

\subsection{\emph{Quiescent Fraction}}
\subsubsection{Dividing by sSFR}
Our first step is to split galaxies into star forming and quiescent
populations. We preferred not to simply divide our population by eye
and sought an automated process which would divide our galaxies in
each redshift bin in a reasonable way. One approach used in the
literature is to divide at a specific star formation rate
$sSFR(z)=1/[3t_{\rm{H}}(z)]$, where $t_{\rm{H}}(z)$ is the Hubble time
at the redshift of interest. This divider in sSFR is roughly
equivalent to the division of galaxies into star forming and quiescent
on the UVJ color-color diagram as described in \citet{Whitaker2012b}
and \citet{Muzzin2013}. This division line, as well as others that we
attempted to use, all shared the same problem: the distribution of
sSFRs in the model and from observations is somewhat different,
especially at z>2.2, so dividing lines which made a reasonable cut for
model galaxies did not work as well for observed galaxies and vice
versa. The sSFR distribution of model galaxies is not as bimodal as it
is for the observations; rather than having a second peak at very low
sSFR, our model distribution tails off. We don't expect this to
significantly affect the results of this work as the star formation
rates in question are already very low (our model galaxies \emph{are}
being quenched; their sSFRs just aren't distributed in quite the same
way as the observations) and any new stars formed shouldn't change the
structural parameters with which we concern ourselves later. Still,
this makes defining quiescence by examining the trough between
populations somewhat difficult. We instead seek to define our
dividing line in relation to the star forming main sequence, which
leads us to a slightly different issue.

The star formation rates of the observed galaxies are systematically
slightly higher than those of the model galaxies, so a typical
observed star forming galaxy (one which we would say occupies the main
sequence of star formation) has a different sSFR than a corresponding
model galaxy. The dependence of sSFR on stellar mass for star forming
galaxies is also steeper for observed galaxies than for model
galaxies. This may point to a deficiency in some of our prescriptions
for star formation and/or stellar feedback \citep[see the discussion
  in][and references therein]{Somerville_Dave:2014}. However, in this
work we are concerned with broadly distinguishing between star forming
and quiescent, and with the processes responsible for moving galaxies
fairly dramatically off of the main sequence. Therefore as long as we
define our dividing line relative to the main sequence in the models
and in the observations, our analysis should be robust.

To deal with these issues, we introduce a method to calculate a
dividing line between star forming and quiescent galaxies which we
apply to both the observations and the model galaxies; however, the
actual normalization, slope, and redshift dependence of the dividing
line are not the same for the model and the observational
samples. \citet{Geha:2012} has shown that, in the local universe,
essentially all isolated galaxies with $m_{\rm star} \lesssim 10^9
M_{\odot}$ show active star formation. This is also the case in our
models. Therefore, at low stellar masses we should be able to measure
the ``native'' star-forming main sequence (SFMS), unaffected by
internal quenching processes. We cannot reliably reach such low mass
limits, but we use galaxies with stellar masses between $10^{9}$ and
$10^{9.5}M_{\odot}$ to measure the baseline SFMS (we restrict our
sample in the models to central galaxies for reasons mentioned in
Section 4.3). We then find the mean log(sSFR) of galaxies in this mass
range in time bins, tracking the evolution of the sSFR of typical star
forming galaxies across cosmic time. Once this evolution is known, we
calculate the main sequence slope by measuring the change in the mean
log(sSFR) between stellar masses of $~10^{9}$ and $10^{10}M_{\odot}$. In a
given redshift bin, we use the mean low-mass sSFR and derived slope to
define a mass-dependent main sequence line. We then define quiescent
galaxies as having less than 25\% of the sSFR of the main sequence
line. Our quiescence divisor for a given redshift and stellar mass is
given by

\begin{eqnarray}
sSFR(z, M_{*})=0.25[10^{MS(z)+ b(log(M_{*}/M_{\odot})-9.25)}]
\end{eqnarray}

where $b$ is the slope we derived and $MS$ is the mean log(sSFR) measured
in the low mass bin
($10^{9}M_{\odot}$$\leq$$M_{*}$$\leq$$10^{9.5}M_{\odot}$). The values
of these quantities are determined separately for the model galaxies
and for the observed galaxies. The coefficients for $MS(z)$ and the
values of $b$ in each of our redshift bins are listed in Tables 1 and
2, respectively.

\begin{table*}
\centering
\caption{Coeffients for $MS(z)=a_{3}t^{3}(z)+a_{2}t^{2}(z)+a_{1}t(z)+a_{0}$, the mean log(sSFR) of the main sequence, where $t(z)$ is the age of the universe in Gyrs at the redshift of interest.}
\begin{tabular}{llccc}
\hline \hline
Dataset & $a_{3}$ & $a_{2}$ & $a_{1}$ & $a_{0}$ \\
\hline \hline
SAM & -0.0012 & 0.039 & -0.499 & -7.640 \\
GAMA \& CANDELS & -0.0017 & 0.039 & -0.398 & -7.513 \\
\hline \hline
\end{tabular}
\end{table*}

\begin{table*}
\centering
\caption{Slope derived as described in the text for each of our redshift bins.}
\begin{tabular}{llcc}
\hline \hline
Redshift & SAM $b$ & GAMA \& CANDELS $b$ \\
\hline \hline
$0.006<z<0.12$ & 0.021 & -0.303 \\
$0.5<z<1.0$ & -0.105 & -0.400 \\
$1.0<z<1.4$ & -0.054 & -0.144 \\
$1.4<z<1.8$ & -0.241 & -0.130 \\
$1.8<z<2.2$ & -0.377 & -0.236 \\
$2.2<z<2.6$ & -0.408 & -0.256 \\
$2.6<z<3.0$ & -0.487 & -0.370 \\
\hline \hline
\end{tabular}
\end{table*}

\begin{figure*}
  \epsfig{file=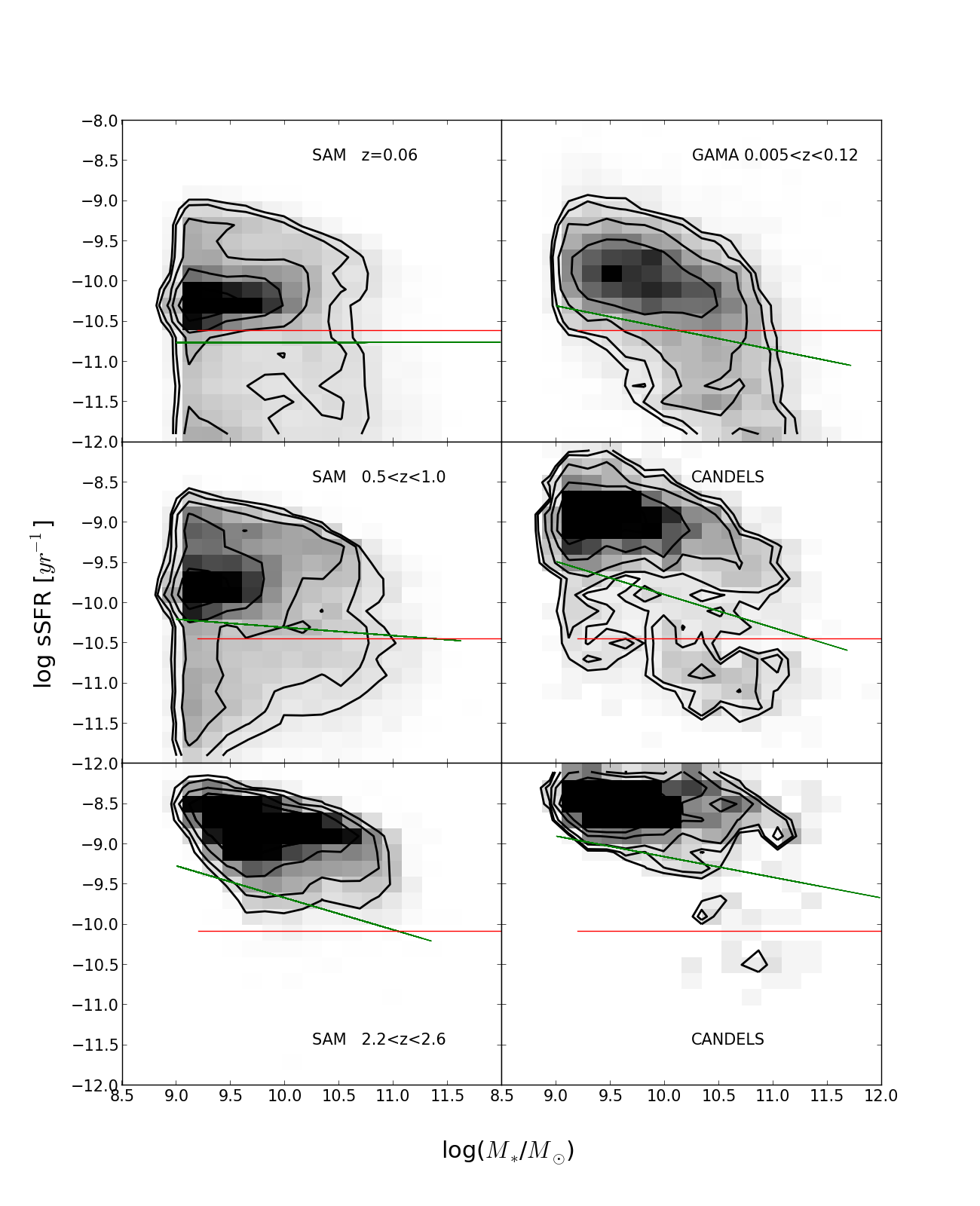, width=1.0\textwidth}
  \caption{The distribution of observed and model galaxies (log($M_{*}$/$M_{\odot}$)>9.0) in the
    plane of stellar mass and specific star formation rate (sSFR) in
    the redshift bins $z\sim0.06$, $0.5<z<1.0$ and $2.2<z<2.6$. The
    greyscale shows the population density with contours overplotted
    in black. The green line shows our adopted dividing line between
    star forming and quiescent galaxies. In practice, the dividing
    line is calculated for each galaxy individually based on its
    stellar mass and redshift; the green line is a least mean squares
    fit to the stellar masses and threshold sSFRs of each galaxy. The
    red line is the $1/3t_{\rm{H}}$ dividing line sometimes used in
    the literature. It is clear that the normalization and slope for
    the model SFMS is different from those for the observations,
    necessitating the use of a different dividing line. Left panel:
    Galaxies from the SAM. Right panel: Galaxies from the GAMA or
    CANDELS survey.}  {\label{splits}}
\end{figure*}

Figure \ref{splits} shows the division of star forming and quiescent
galaxies for the model (including disk instability) and from the
observations in the redshift bins $z\sim0.06$, $0.5<z<1.0$ and
$2.2<z<2.6$. Although the galaxy sample we will use for the remainder
of this work includes only galaxies with
log($M_{*}$/$M_{\odot}$)>10.0, here we plot galaxies down to stellar
masses of $10^{9} \, M_{\odot}$, since these are the galaxies from
which our dividing lines are derived. The green line indicates our
split between star forming and quiescent galaxies. The red line is
drawn at $sSFR=1/(3t_{\rm{H}})$, where $t_{\rm{H}}$ is the Hubble time
at the median redshift of the bin, for comparison with alternate
dividing lines commonly used in the literature. We apply the cut
derived for the DI model to the noDI model as well, since we would
like to see how the disk instability affects the sSFRs of galaxies
within the model and that information would be lost if we allowed the
cut to move between the two models. It is worth noting, however, that
$MS(z)$ and $b$ are very similar between the two models. We can see in
all three bins that the dividing line has a different slope and
normalization for the models than for the observations.

\begin{figure}
  \epsfig{file=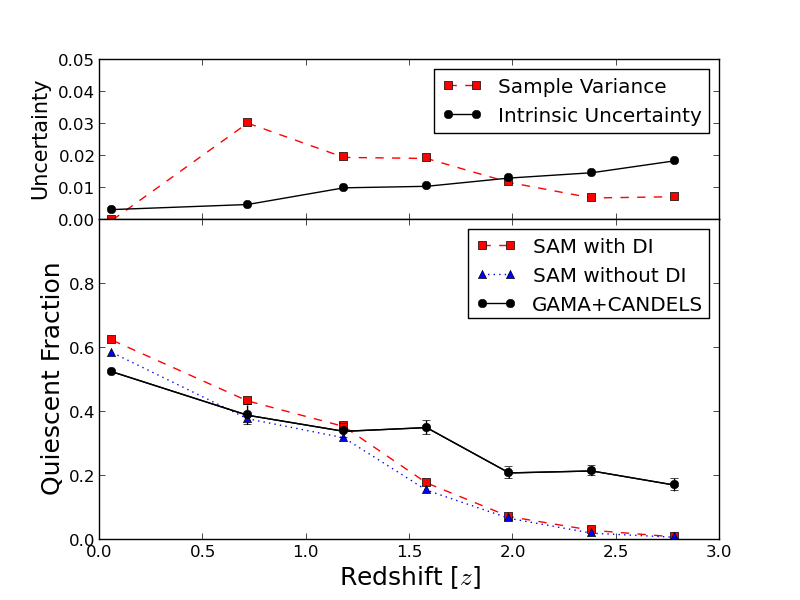, width=0.5\textwidth}
  \caption{The evolution of the quiescent fraction of galaxies
    (log($M_{*}$/$M_{\odot}$)>10.0) with redshift. The top panel is
    the predicted 1-$\sigma$ uncertainty due to sample variance (red,
    dashed) and due to uncertainty in galaxy parameter estimation
    (black, solid). These are added together in quadrature and shown
    plotted on the observational measurements in the bottom panel. In
    the bottom panel the red dashed line with squares corresponds to
    the model including disk instabilities, the blue dotted line with
    triangles to the model without disk instabilities, and the black
    solid line with circles to the observations. This convention is
    used throughout this work. Field-to-field variance is expected to
    be negligible in the lowest redshift bin, so here the plotted
    error is entirely due to uncertainties in galaxy
    properties. Overall, the agreement between the model predictions
    and observational results is quite good. Below $z\sim1.2$, the
      fractions predicted by the model differ from the observational
      fraction by no more than 0.1, although at $z\gtrsim1.2$ they
      begin to differ by as much as 0.2, with the model predicting
      almost no quiescent galaxies. The predicted quiescent fraction
    is affected very little by the inclusion of disk instabilities in
    the model.}  {\label{qfrac}}
\end{figure}

\subsubsection{Evolution of the Quiescent Fraction}

Figure \ref{qfrac} shows the evolution of the quiescent fraction of
galaxies with redshift for galaxies from the SAM, both the DI and noDI
models, and from observational data taken from GAMA and CANDELS. We
compute 1-$\sigma$ uncertainties due to field-to-field variance and
uncertainty in observed galaxy properties (stellar mass, S{\'e}rsic
index and star formation rate) as follows. Our lightcones are about
nine times larger than the CANDELS fields that we are comparing with,
so we select a model sample from a subsection of the lightcone that
has comparable area. If we select different CANDELS-sized areas from
our lightcone to do our analysis, we get a measure of the effect of
cosmic variance. We also calculate the 1-$\sigma$ error in the
quiescent fraction due to uncertainties in the estimates of galaxy
properties in the observational sample. We use quoted uncertainties in
S{\'e}rsic index, assume an uncertainty of 0.25 dex for star formation
rates and use the redshift-dependent stellar mass uncertainty of
\citet{Behroozi2013b}. The separate uncertainties due to cosmic
variance and parameter estimation can be seen in the top panel. We add
the uncertainty due to each in quadrature and apply them to the
observations. In the lowest redshift bin, the error estimates reflect
only the uncertainties due to errors in the physical parameters; these
uncertainties dominate over the cosmic variance due to the large
volume probed by GAMA. We note here that we are still likely
underestimating uncertainties due to systematics such as the assumed
star formation histories of CANDELS galaxies, possible variations in
the IMF, etc.

The quiescent fraction of galaxies in the model is relatively
insensitive (changing by <10\% in all redshift bins) to the inclusion
of disk instabilities in our models; as we will see, the net effect of
the disk instability is mainly to create more bulge-dominated
galaxies. This is due in part to the fact that our disk instability
prescription does not affect gas and limits the amount of low-level
AGN feedback that is triggered by disk instabilities. Both models
agree well with observations at low redshift; for $z\lesssim1.2$, the
fractions differ by no more than $0.05-0.1$. Above this redshift,
however, the fractions begin to differ by about $0.2$, with the model
predicting fewer quiescent galaxies than are observed. Overall, the
model exhibits a steeper evolution than the observed galaxies,
predicting basically no quiescent galaxies at $z\sim3$. It seems that
the model is not quenching galaxies early enough. We will discuss
possible reasons for this discrepancy later.

\begin{figure*}
  \epsfig{file=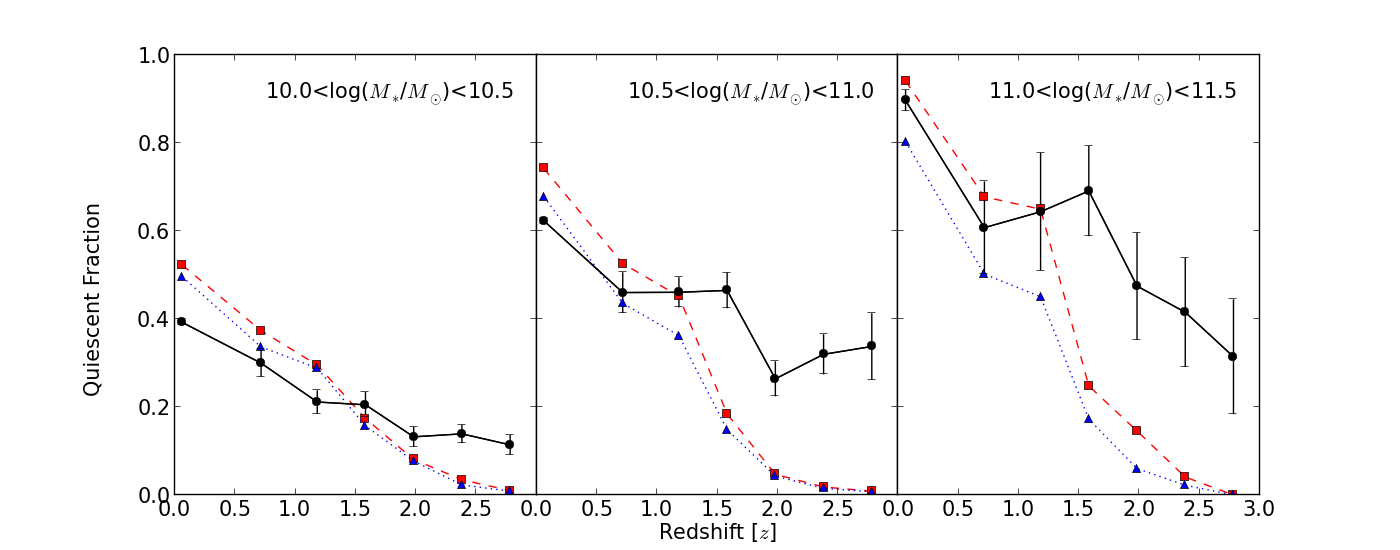, width=1.0\textwidth}
\caption{Each panel is like Figure \ref{qfrac}, but for different bins
  in stellar mass. Line types and colors are the same as Figure
  \ref{qfrac}. The quiescent fraction increases with stellar mass for
  both the models and observations, with disk instabilities
  contributing more of the quenched galaxies at higher masses. The
  discrepancy between the model predictions and observations is larger
  in the two higher stellar mass bins. } {\label{qfracmass}}
\end{figure*}

In Figure \ref{qfracmass}, we examine the mass dependence of the
quiescent fraction evolution.  The behavior is similar in each mass
bin to the overall behavior in Figure \ref{qfrac}. At redshifts above
$\sim1.2$ the model predicts a smaller quiescent fraction than is
observed for all three mass bins. This discrepancy gets worse as the
stellar mass increases. In the highest mass bin, the fractions can
differ by as much as 40 \%. This is an extension of the overall high
redshift discrepancy in Figure \ref{qfrac}; since our model predicts
no quiescent galaxies (in any mass range) and in general more massive
galaxies are likely to be considered quiescent, the gulf between our
model quiescent fraction and the observed fraction widens as the
masses considered become larger. In the two lower mass bins,
  the DI model actually overproduces quiescent galaxies by as much as
  0.15 at $z\lesssim1.2$, but it is in better agreement (within 0.05) with
  observations in the highest mass bin. As expected, the quiescent
fraction increases for galaxies with higher stellar mass for both the
model and the observations.  The model also captures the steeper
evolution of the quiescent fraction for higher masses, although again,
the evolution in the model is steeper than observations for all mass
bins.

\begin{figure}
  \epsfig{file=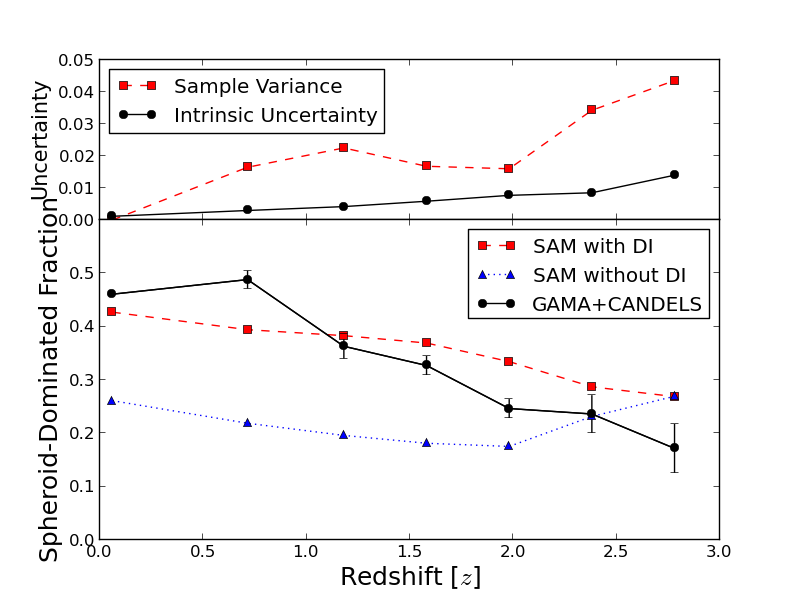, width=0.5\textwidth}
  \caption{The evolution of the spheroid-dominated fraction of
    galaxies with redshift.  Error bars are the $1-\sigma$
    uncertainties due to sample variance and uncertainties in observed
    galaxy properties added in quadrature, as in Fig.~\ref{qfrac}. The
    separate contributions are plotted in the top panel. The model in
    which spheroids form only via mergers underproduces the fraction
    of spheroid dominated galaxies at $z \lesssim 2$ and does not
    reproduce the build-up of the spheroid-dominated population seen
    in the observations. The model with additional spheroid growth via
    disk instabilities (DI model) is qualitatively in fairly good
    agreement with observations (the two agreeing to within
      $\sim0.1$ at all redshifts), though the predicted evolution in
    spheroid fraction is still a bit too shallow, with the model
    overpredicting spheroid-dominated galaxies at high redshift and
    underpredicting them at low redshift.  } {\label{efrac}}
\end{figure}

\subsection{\emph{Spheroid-Dominated Fraction}}

We now split galaxies into spheroid-dominated and disk-dominated
populations. We define spheroid-dominated galaxies as having
S{\'e}rsic indices greater than 2.5, the average of a pure disk
($n=1$) and pure bulge ($n=4$), as has been done in many other studies
\citep{Shen2003, Lange2014, Bruce2014, Mortlock2015}. We discuss later
how making this spheroid-domination cut less stringent (at $n=2$)
affects our results. We have also done the same analysis by dividing
galaxies at a bulge-to-total mass ratio of 0.5. These results are very
similar and can be found in Appendix B.  Figure \ref{efrac} shows the
evolution with redshift of the spheroid-dominated fraction of
galaxies. Here, we see the main effect of the disk instability. The
noDI model severely underpredicts the fraction of spheroid-dominated
galaxies at almost all redshifts, with the disagreement becoming worse
towards lower redshifts. The DI model does a much better job of
matching the observed spheroid-dominated fraction, increasing our
prediction by almost a factor of two at low redshift; within the
uncertainties, the fractions do not disagree by more than $0.1$ at any
redshift. However, the evolution of the spheroid-dominated fraction in
the model is somewhat shallower than in observations, so we
overpredict the spheroid-dominated fraction at $z\gtrsim1$ and
underpredict it at $z\lesssim1$.

\begin{figure*}
  \epsfig{file=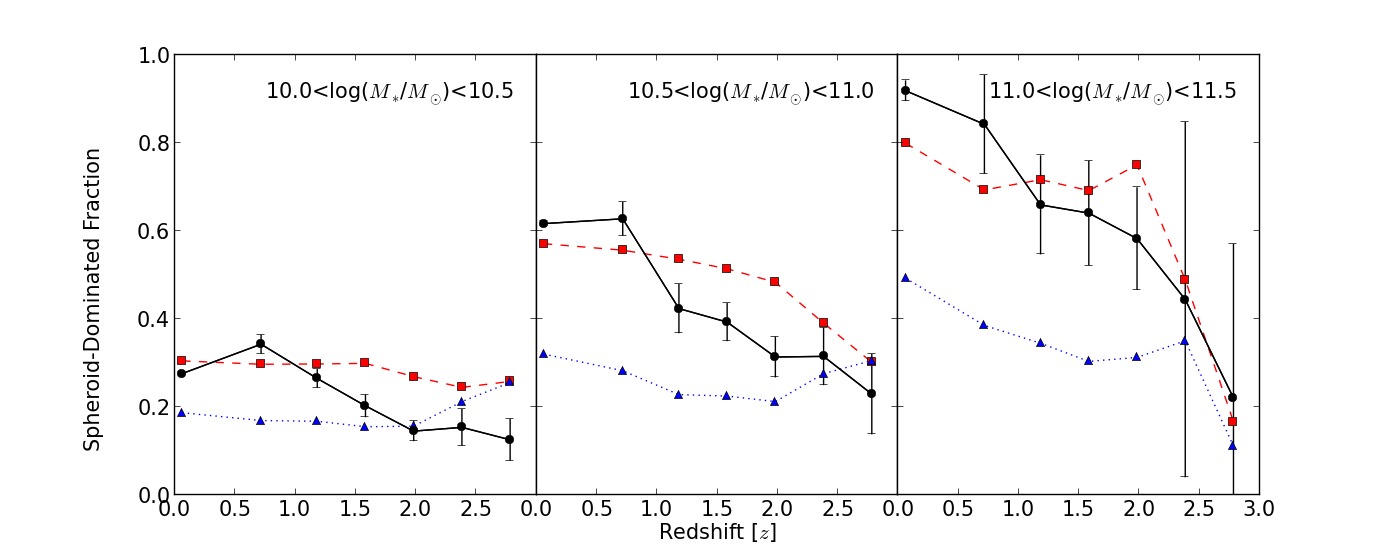,
    width=1.0\textwidth}
  \caption{Same as Figure \ref{qfracmass}, but for the spheroid-dominated fraction.}
  {\label{efracmass}}
\end{figure*}

We once again investigate the mass dependence in Figure
\ref{efracmass}. Again, as expected, at larger masses, the
spheroid-dominated fraction is greater at almost all redshifts, and
the spheroid-dominated fraction increases more rapidly with redshift
for massive galaxies. There appears to be a population of massive,
disk-dominated galaxies at high redshift in both the observations and
the model. The behavior in all three mass bins is reminiscent of the
overall behavior in Figure \ref{efrac}, except for the high mass, high
redshift case. As in Figure \ref{efrac}, the disk instability brings
the model mainly into better agreement, although in the two lower mass
bins at $z\gtrsim1$ the fractions can differ by as much as $0.15-0.2$,
which is slightly more than for the overall population.

\subsection{Comparison with Previous Results}

We take a moment here to compare with previous work that has probed
the evolution of the quiescent and spheroid-dominated fractions of
galaxies. \citet{Brammer2011} and \citet{Muzzin2013} both examine the
dependence of the quiescent fraction on stellar mass across a range of
redshifts. \citet{Brammer2011} examine galaxies from the NEWFIRM
Medium-Band Survey (NMBS) \citep{vanDokkum2009, Whitaker2010} while
\citet{Muzzin2013} observe a sample of galaxies in the
COSMOS/UltraVISTA field \citep{Muzzin2013b} over a similar redshift
range. Our observational results are in good agreement with both
studies. The quiescent fraction is higher at larger stellar masses and
lower redshifts and the quiescent fraction of high stellar mass
galaxies increases more steeply with redshift than that of low mass
galaxies. The quiescent fraction evolution of our high mass bin
($10^{11} M_{\odot} \lesssim M_{*} \lesssim 10^{11.5} M_{\odot}$) is
in good agreement with \citet{Brammer2011}; in both cases, the
fraction increases from $\sim 0.5$ at $z\sim2$ to $\sim0.8$--0.9 at
$z\sim$ 0.5. \citet{Muzzin2013} is in good agreement with our high
mass quiescent fraction as well, but also investigates the quiescent
fraction down to lower stellar mass so we can compare our lower mass
bins. These also agree very well. For galaxies with $10^{10} M_{\odot}
\lesssim M_{*} \lesssim 10^{10.5} M_{\odot}$, the quiescent fraction
increases from $\sim 0.2$ at $z\sim$ 2 to $\sim 0.4$ at $z\sim$
0.2--0.5. Meanwhile, the quiescent fraction of galaxies with
$10^{10.5} M_{\odot} \lesssim M_{*} \lesssim 10^{11} M_{\odot}$
increases from $\sim0.4$ at $z\sim2$ to $\sim0.5$--0.6 at $z\sim$
0.2--0.5.

In terms of spheroid-dominated fraction, \citet{Buitrago2013} examines
the fraction of spheroid-dominated galaxies with masses
>$10^{11}M_{\odot}$ from $z\sim3$ to the present day. They cover this
range by combining several different surveys: SDSS DR7
\citep{Abazajian2009}, POWIR/DEEP2 \citep{Bundy2006, Conselice2007}
and the GOODS NICMOS Survey (GNS;\citealt{Conselice2011}). They find a
steady increase in the fraction of spheroid-dominated galaxies
($n>2.5$) whereas, when the same mass cut is applied (as can be seen
in the right panel of Figure \ref{efracmass}), we predict a sharper
increase in spheroid-dominated fraction from $\sim20$\% at $z\sim2.5$
to $\sim60$\% at $z\sim1.5$. Because of this, we predict a somewhat
larger spheroid-dominated fraction than theirs between $z\sim1$ and
2. \citet{Bruce2014} observe galaxies with $M_{*}$>$10^{11}M_{\odot}$
in the COSMOS and UDS fields from the CANDELS survey over the redshift
range 1<$z$<3. They use bulge-disk decompositions to sort galaxies by
$B/T$ and compute the spheroid-dominated fraction. They find a
spheroid-dominated fraction of $\sim0.6$ for $z\sim1.5$ and a fraction
of $\sim0.45$ for $z\sim2.5$. This is in very good agreement with our
results (again cutting at $10^{11}M_{\odot}$ and now defining
spheroid-dominated as having $B/T$>0.5); we find a spheroid-dominated
fraction of $\sim0.45$ for $z\sim2.5$ and $\sim0.65$ for $z\sim1.5$.

\subsection{\emph{Dividing into Quadrants}}

\begin{figure*}
  \epsfig{file=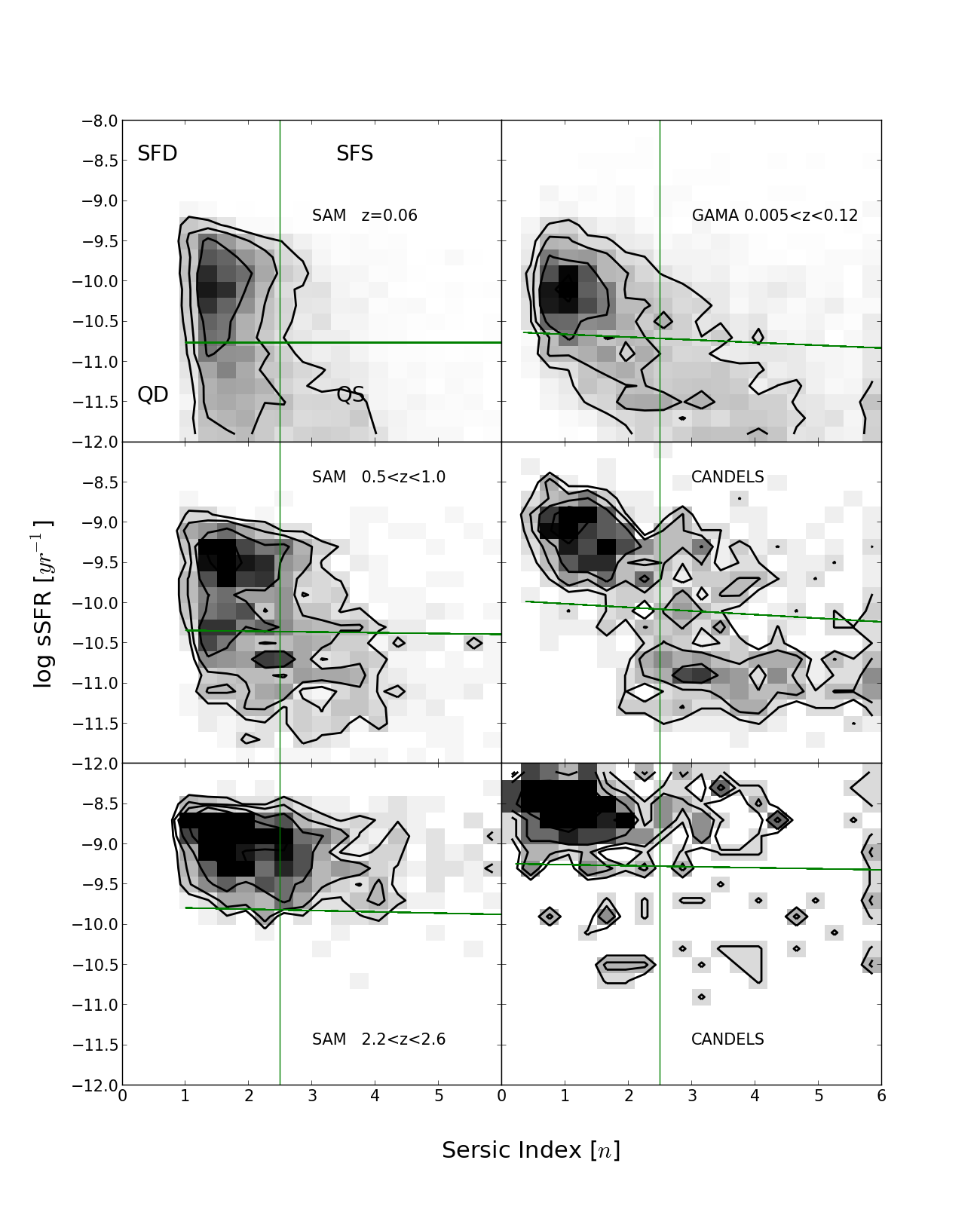, width=1.0\textwidth}
  \caption{The distribution of galaxies in the sSFR-$n$ plane in the
    redshift bins $z\sim0.06$, $0.5<z<1.0$ and $2.2<z<2.6$. Left
    panel: Galaxies from the SAM. Right panel: Galaxies from GAMA and
    CANDELS. The greyscale shows the population density in the
    sSFR-$n$ plane, with contours in black overplotted. The green
    lines are the dividing lines used in this work to identify the
    four ``quadrants'' (see text).}  
    {\label{quadsplits}}
\end{figure*}

\begin{figure*}
  \epsfig{file=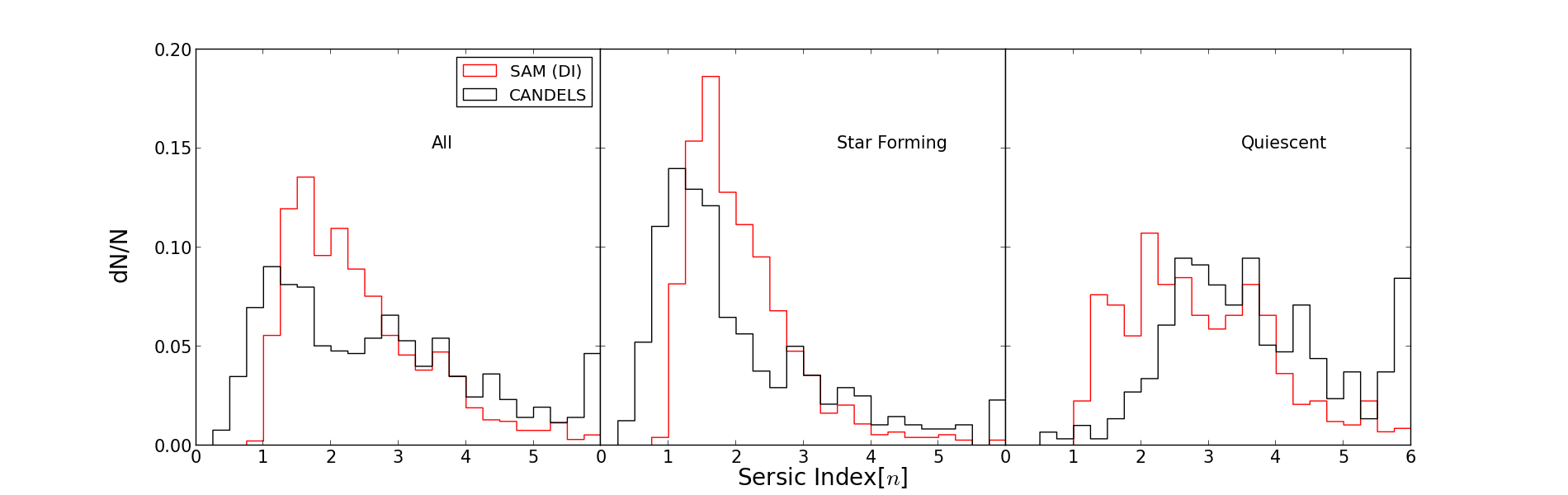, width=1.0\textwidth}
\caption{Distribution of S{\'e}rsic indices for model galaxies and
  CANDELS galaxies in the redshift bin 0.5<z<1.0. Left panel: All
  galaxies. Middle Panel: Star forming galaxies. Right panel:
  Quiescent galaxies.}  {\label{ndist}}
\end{figure*}

\begin{figure*}
  \epsfig{file=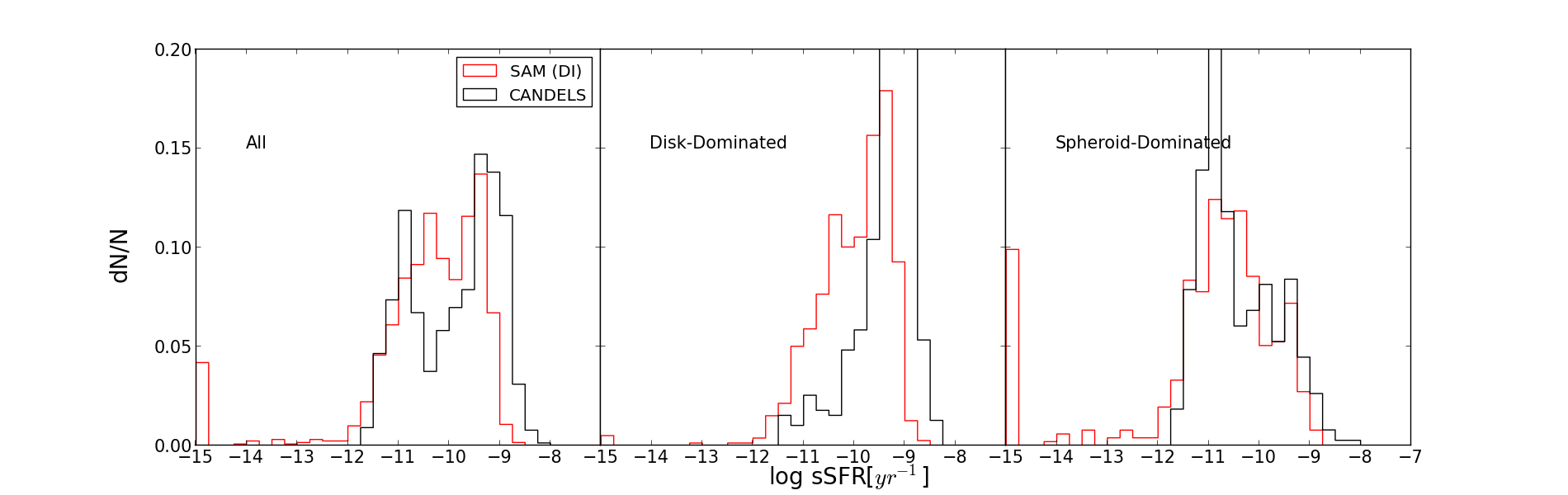, width=1.0\textwidth}
  \caption{Same as previous figure, but now showing the distribution
    of sSFR. Left panel: All galaxies. Middle Panel: Disk-dominated
    galaxies. Right panel: Spheroid-dominated galaxies. } {\label{ssfrdist}}
\end{figure*}

Having divided the sSFR-S{\'e}rsic plane in halves, we now further
divide the plane into four quadrants to examine the evolution of the
populations in each one: SFDs, QSs, SFSs and QDs. Figure
\ref{quadsplits} shows an example of the division of galaxies into
quadrants for both the model and the observations in the redshift bins
$z\sim0.06$, $0.5<z<1.0$ and $2.2<z<2.6$. The star formation division
line is a least mean squares fit to the individual star formation
thresholds for each galaxy in the redshift bin according to its
stellar mass and specific redshift. Figures \ref{ndist} and
\ref{ssfrdist} show the distributions of S{\'e}rsic index and sSFR for
our DI models and for the observations in the redshift bin
$0.5<z<1.0$. In both cases, the distributions are similar, but not
exactly the same.  Our model has trouble reproducing the strong
observed bimodality in both quantities; our disk instability creates
many galaxies of intermediate S{\'e}rsic index.  As we move toward
lower redshift, the differences between the distributions of model and
observed galaxies become more significant. We will return to this
point in the discussion.

\begin{figure*}
  \epsfig{file=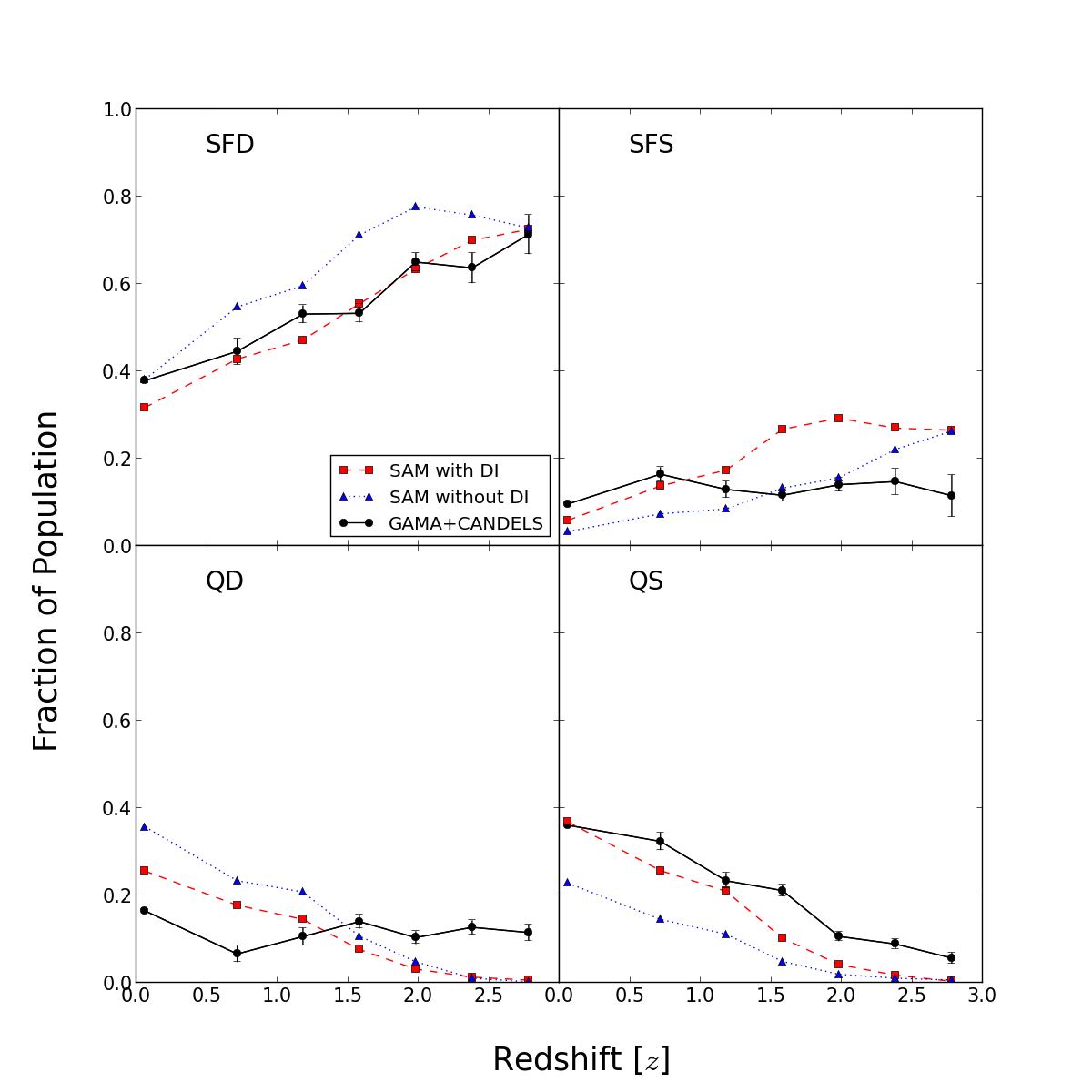, width=1.0\textwidth}
  \caption{The evolution of the fraction of galaxies in each quadrant
    of the sSFR-S{\'e}rsic plane with redshift. Top left: Star forming
    disk-dominated galaxies (SFD). Top right: Star forming
    spheroid-dominated galaxies (SFS). Bottom left: Quiescent
    disk-dominated galaxies (QD). Bottom right: Quiescent
    spheroid-dominated galaxies (QS). Our models qualitatively
    reproduce the trends of a decreasing fraction of SFD galaxies and
    the increasing fraction of QS galaxies with cosmic time, with the
    DI model in general producing better agreement with the
    observations. Our models do less well at reproducing the observed
    trends for SFS and QD, predicting mild decrease and increase in
    these populations, respectively, with cosmic time, while in the
    observations their fractions are nearly constant from $3 \gtrsim z
    \gtrsim 0.1$.  } {\label{quadevol}}
\end{figure*}

Figure \ref{quadevol} shows the evolution of the fraction of all
galaxies in each quadrant with redshift for the DI model, the noDI
model and the observations. We see here again the reason for the
difference in how the quiescent and spheroid-dominated fractions
change with the disk instability: the disk instability decreases the
fraction of QDs while increasing the fraction of QSs, leaving the
quiescent fraction relatively unchanged. The two spheroid-dominated
populations, however, are both increased, leading to the large change
in the overall spheroid-dominated fraction. The DI model reproduces
the evolution of SFDs to within a few percent as their
numbers dwindle due to various transformative processes. The noDI
model predicts too many SFDs, the model and observed fractions
  differing by as much as 0.2. The DI model reproduces the observed
fraction of QSs at $z\sim 0.1$, but slightly underproduces QSs at
higher redshifts, although the fractions do not differ by more than
$\sim0.1$. The noDI model underproduces QSs at all redshifts, to an
even larger degree. Both models underpredict the fraction of QDs at
$z\gtrsim1.5$ (which is again an extension of the overall issue seen
in Figure \ref{qfrac}) and overpredict them at $z\lesssim1.5$. Once
again, the disagreement is worse when the model with no disk
instability is considered. While the models match the observed
fraction of SFSs to a few percent at redshifts $\lesssim1$, they
predict too many at high redshift, in some cases by a factor of
two. It becomes clear when comparing the two models that the disk
instability is mostly responsible for the excess of SFSs that we
predict at redshifts $z\sim1.5-2.5$. At redshifts higher than this,
mergers seem to become increasingly important as a channel for bulge
growth. We expect some SFSs in the universe to have disturbed
morphologies due to the process responsible for making them an SFS. It
is possible that some of the CANDELS galaxies that would be classified
as SFSs are dust obscured and are either not detected or not
considered star forming, leading to an underestimate of the fraction
of SFSs. While we do include the effect of dust extinction in our
model, as well as make the same $H$-band magnitude cut as is used for
CANDELS, the possibility remains that we are underestimating dust
extinction. This would cause objects that are missed in CANDELS due to
the $H$-band magnitude limit to be included in our model catalogs. It
is not unreasonable that we would be underestimating the effects of
dust in these objects in particular, as our prescription is based on
an undisturbed disk geometry and does not account for the possibly
heavily-obscured starbursting systems we are concerned with in the SFS
quadrant.

\begin{figure*}
  \epsfig{file=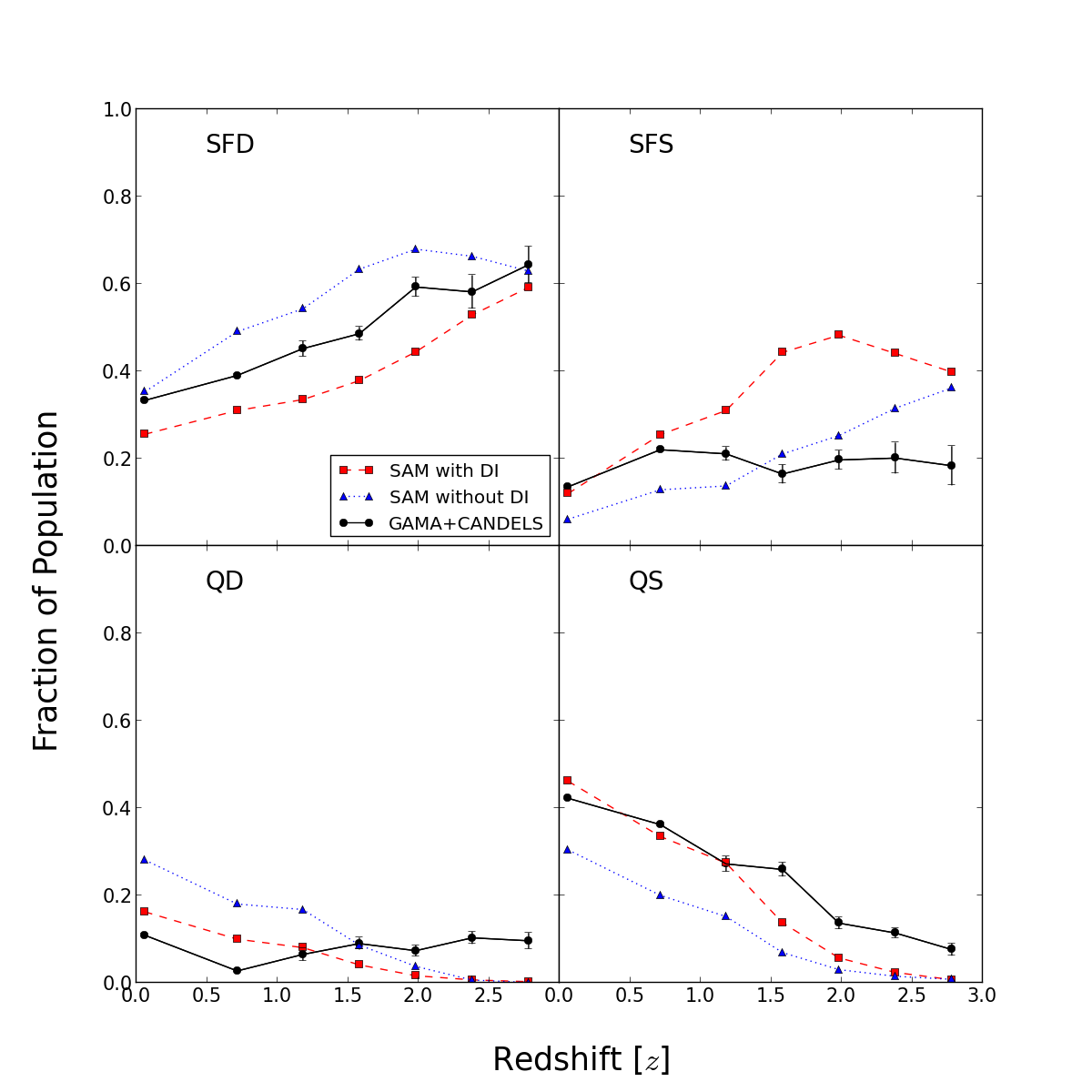, width=1.0\textwidth}
  \caption{Same as Figure \ref{quadevol}, but now with a morphology cut at $n=2$.} {\label{quadevol2}}
\end{figure*}

We also note here that changing our cut in S{\'e}rsic index from
$n=2.5$ to $n=2$, which still distinguishes systems with significant
bulge components, does change our results somewhat as the distribution
of $n$ in the models is different from the observed distribution (as
seen in Figures \ref{quadsplits} and \ref{ndist}). The spheroid-dominated fraction
increases more for the DI model than for the observations, especially
at higher redshifts. The noDI model is changed very little. When
looking at different mass bins as in Figure \ref{efracmass}, the
change of the spheroid-dominated fraction of the DI model relative to
the observations is more pronounced in the two lower mass bins than in
the highest one. Figure \ref{quadevol2} shows the evolution of the
fraction of all galaxies in each quadrant for the morphology cut at
$n=2$. The DI model now underpredicts the fraction of SFDs by about 0.1-0.15 at all redshifts and
overpredicts SFSs at high redshift by an even larger amount (as much as 0.4). The
fraction of QSs matches the observational results well at $z \lesssim
1.5$ but still underpredicts these objects at higher
redshifts. However, qualitatively the results are very similar, so we
continue to use our $n=2.5$ cut for the rest of this work.

Now, knowing both where our model succeeds and fails in matching the
buildup of these populations, we can dig into the model to see which
mechanisms are responsible for moving our simulated galaxies in the
sSFR-$n$ plane.

\section{Discussion}
The SAM can provide us with details about galaxy formation histories
which we cannot glean directly from observations; we now examine the
statistics of events that drive galaxy tranformation and quenching
(mergers and disk instabilities) in our models, and provide
representative examples of how individual galaxies trace out their
histories in the sSFR-$n$ plane.

\begin{figure*}
  \epsfig{file=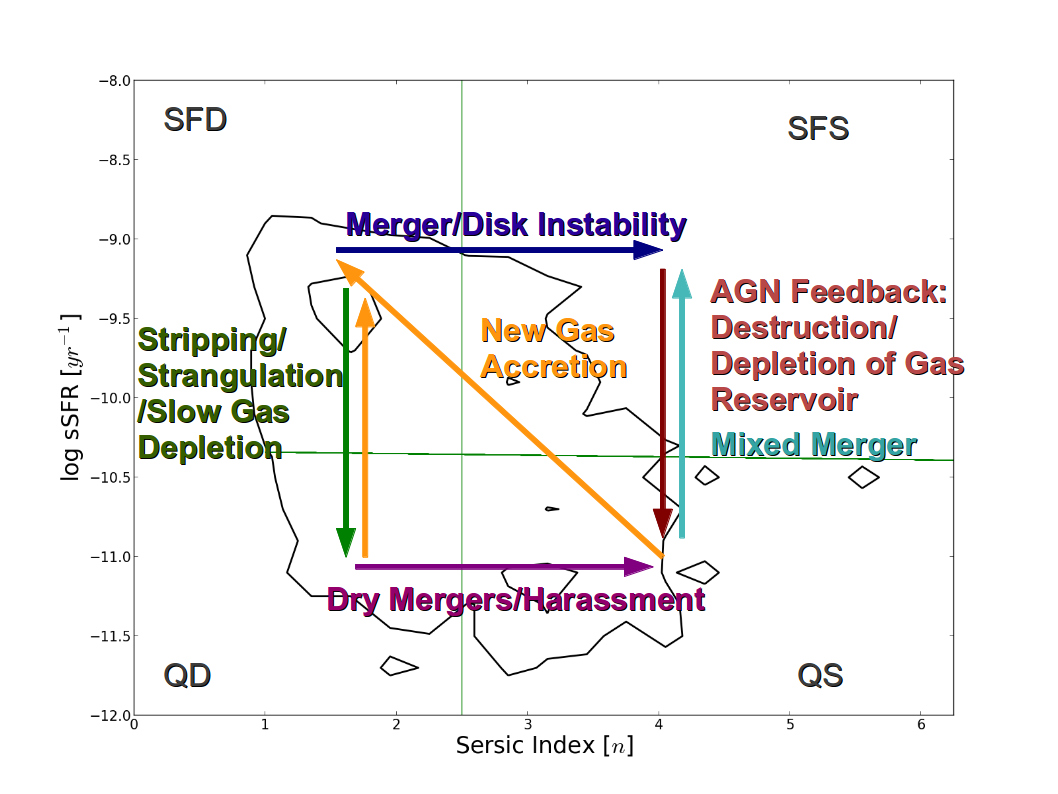, width=1.0\textwidth}
  \caption{Schematic representation of how different physical processes
  might cause galaxies to migrate in the sSFR-$n$ plane. The density
  distribution of galaxies in the SAM (DI model) for $0.5<z<1.0$ is shown with
  contours.}  
  {\label{quad}}
\end{figure*}

Figure \ref{quad} shows density contours for galaxies from the SAM
($0.5<z<1.0$) in the sSFR-$n$ plane. Overlaid arrows show how
different physical processes might move galaxies in this diagram. SFDs
may merge with each other or suffer disk instabilities to form
bulge-dominated galaxies which then undergo gas depletion by AGN
feedback, leading to quenching of star formation. In the SAM this
occurs over relatively short time scales of several hundred million
years. Meanwhile, other SFDs may passively evolve, depleting their gas
reservoirs over much longer time scales of a Gyr or more, eventually
becoming QDs, which may then experience dry (gas-poor) mergers which
puff them up and form QSs. Quiescent galaxies may then accrete new gas
which allows regrowth of a disk component. We now examine the
importance of some of these processes in the SAM in a bit more detail.

\subsection{\emph{How Recently Have Different Types Been Disturbed?}}

\begin{figure*}
  \epsfig{file=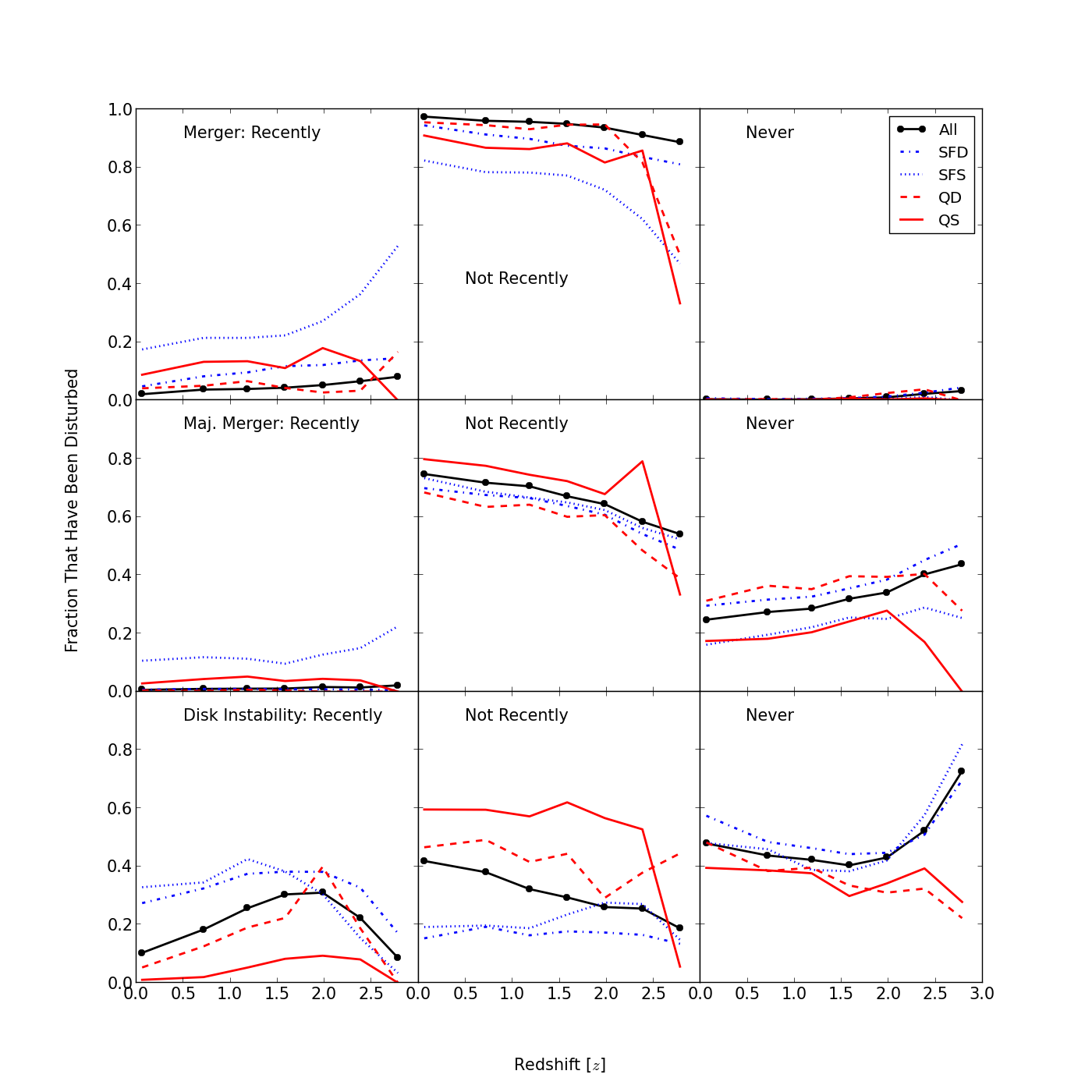, width=1.0\textwidth}
\caption{Top row: Left panel: Fraction of model galaxies in each
  quadrant which have undergone a recent (<$3t_{\rm{dyn}}$) merger with
  mass ratio >1:10. Middle panel: Fraction which have undergone a
  merger on a timescale >$3t_{\rm{dyn}}$. Right panel: Fraction which
  have never undergone a merger. Middle row: Same as top row, but for
  major mergers (>1:3). Bottom Row: Same as top and middle rows, but
  for disk instabilities. The fraction of galaxies that have suffered
  a recent merger declines with cosmic time from $z\sim 2$ to the
  present, while the fraction that have experienced a recent disk
  instability peaks at around $z\sim 1.5$--2. } {\label{tmfracevol}}
\end{figure*}

We would like to know how galaxies in different quadrants in the
sSFR-S{\'e}rsic plane are formed or evolve to their current state and
with the SAM we can directly measure the time since traumatic events
such as mergers and disk instabilities. In the top left panel of
Figure \ref{tmfracevol}, we look at the fraction of
>$10^{10}M_{\odot}$ galaxies of each type which have undergone a
recent merger, with ``recently'' being defined as within three
dynamical times (where
$t_{\rm{dyn}}=2$$\pi$$r_{\rm{disk}}/v_{\rm{disk}}$). We see that SFSs
are more likely to have experienced a recent merger at all redshifts,
while very few QDs have undergone a recent merger. SFDs and QSs fall
in between. All types of galaxies are more likely to have experienced
a recent merger at higher redshift. In the top right panel, we see
that almost no galaxies at any redshift have avoided ever having a
merger in their lifetime.

In the middle row, we restrict our attention to major mergers with a
mass ratio >1:3 (the mass used to calculate this ratio is the combined
baryonic (cold gas+stellar) and dark matter mass within 2 halo scale
radii; see S08) and see that basically no QDs have undergone recent
major mergers, while the fraction of SFDs with recent major mergers is
only slightly higher (no more than $\sim5$\%). This is because it is
very unlikely to experience a major merger and still retain enough of
a disk to be considered disk-dominated within a dynamical time of the
merger. QSs are slightly more likely to have undergone a recent major
merger than SFDs, while 65-80\% (depending on redshift) have undergone
a non-recent major merger. This is simply because very soon after a
major merger, the merger-triggered starburst would cause the
spheroid-dominated galaxy to be classified as an SFS. After some time
has passed and star formation has been quenched, it would be
classified as a QS. For this same reason, SFSs are the least likely to
have undergone a non-recent major merger; if the major merger wasn't
recent, they're unlikely to still be star forming. SFSs are still most
likely to have had a recent major merger and that likelihood increases
somewhat towards higher redshift. The disk-dominated classes are
actually more likely to have had a non-recent major merger than never
to have had one at all, after which they must have regrown a
substantial disk component. Unsurprisingly, the disk-dominated
populations are more likely to have never had a major merger than the
spheroid-dominated populations.

In the bottom row, we see that the fraction of all types which have
undergone a recent disk instability peaks at $z\sim1.5$. We note that
this is in qualitative agreement with the peak of the clumpy fraction
of galaxies (in the mass range 9.8<log($M_{*}$/$M_{\odot}$)<10.6)
found in \citet{Guo2014}. We cannot make a quantitative comparison
with these results as we have no way to estimate clumpiness in our
models, but clumpy galaxies are expected to be associated with minor
mergers and disk instabilities \citep{Dekel2009}.

Our star forming classes are more likely than our quiescent classes to
have undergone a recent disk instability, but the fraction of SFSs
plummets towards higher redshift, signaling the fact that mergers seem
to be the dominant bulge-growing channel at high redshift, with disk
instabilities increasing in importance as mergers become less
frequent, as has been speculated in previous works
\citep{Parry2009,DeLucia2011,Fontanot2012}. QSs are much more likely
to have never had a disk instability, or to have had one longer ago
(since it has likely been a while since they had a disk). QDs are just
as likely to have had a disk instability recently, not recently, or
not at all, suggesting that disk instabilities don't play a huge role
in their evolution. Finally, a significant fraction of all types
(40-60\%) have never experienced a disk instability. This fraction
increases steeply towards high redshift, presumably because the merger
rate increases and it is less likely for a galaxy to go undisturbed
long enough to develop an instability on its own.

We note that there is a large amount of uncertainty in the highest redshift bin for our quiescent classes. There are very few galaxies classified as quiescent and so we suffer from small number statistics in that bin.

\subsection{\emph{Individual Galaxy Histories}}

To illustrate how individual galaxies evolve, we now inspect the
evolutionary tracks of four galaxies selected from the SAM, which end
up in the four different quadrants of the sSFR-$n$ plane at
$z=0$. Here we use bulge-to-total mass ratio as our proxy for
morphology, since the tracks in the sSFR-morphology plane are much
easier to see this way (and because the results are very similar: see
Appendix B). Figure \ref{sfdtime} shows the evolutionary path of a
galaxy with a fairly quiet history that ends up as an SFD. This galaxy has a mass of $\sim10^{10.8}M_{\odot}$ at $z=0$. The top
panel is its track in the sSFR-morphology plane, color coded by the
age of the universe. The bottom and middle panels are the evolution
with time of the morphology and sSFR respectively. We see that this
galaxy has a few mergers early on after which its evolution is
entirely due to accretion of new material, allowing it to continue
forming stars. The decrease in $B/T$ following the merger events is due
to the regrowth of a disk component. 

\begin{figure}
\epsfig{file=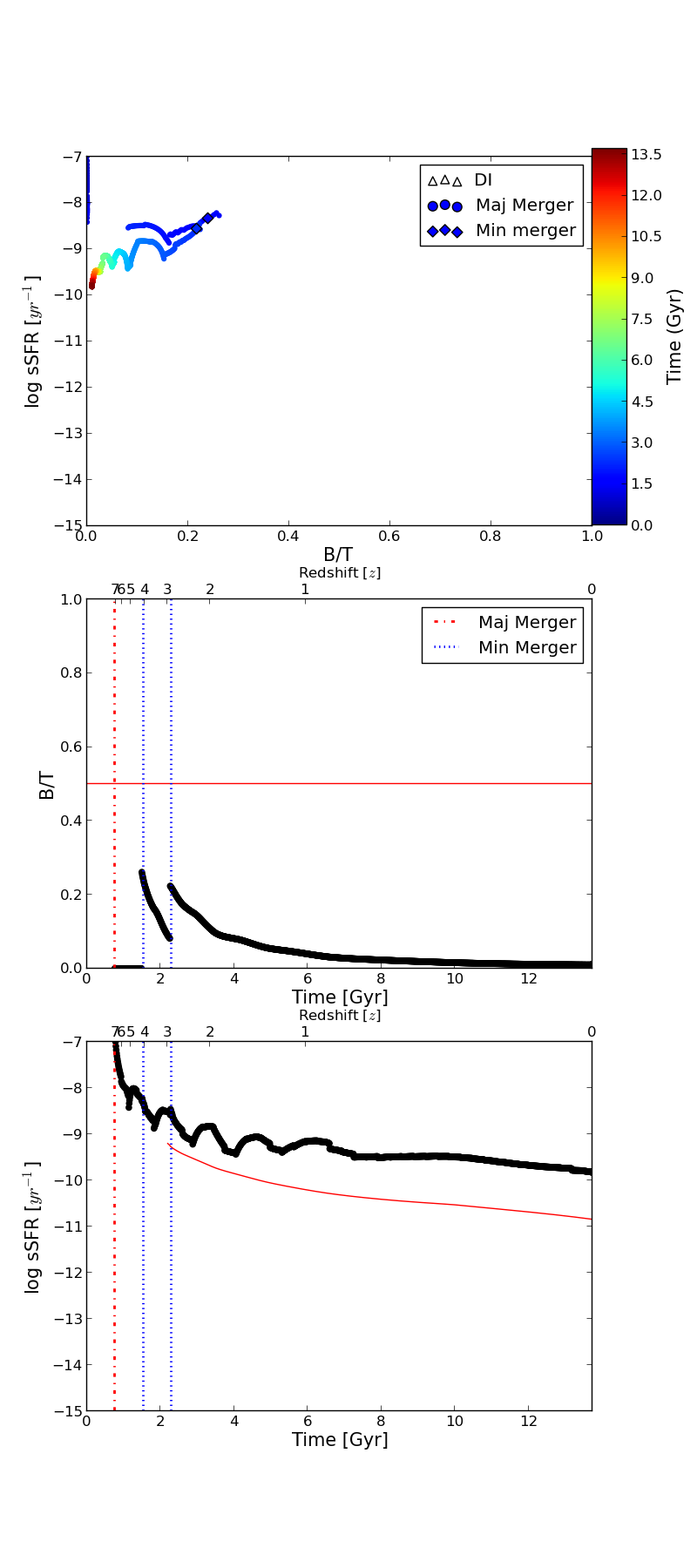,
  width=0.5\textwidth}
\caption{Evolution of sSFR and $B/T$ mass ratio for a galaxy which is
  classified as a star forming disk-dominated galaxy with a
    mass of $\sim10^{10.8}M_{\odot}$ at $z=0$. Top panel:
  Evolutionary track in the sSFR-$B/T$ plane, color coded by age of
  the universe. Minor mergers (<1:3), major mergers (>1:3) and DIs are
  indicated by diamonds, circles and triangles, respectively. Middle
  panel: Evolution of $B/T$ mass ratio with time. The red dash-dotted
  line indicates a major merger and the blue dotted lines indicate
  minor mergers. The solid red line is our division between
  disk-dominated and spheroid-dominated. Bottom panel: Evolution of
  sSFR with time. The solid red line is our division between star
  forming and quiescent. The galaxy remains disk dominated, due to its
  quiet accretion history, and the SFR gradually declines due to the
  declining cosmological accretion rate. } {\label{sfdtime}}
\end{figure}

\begin{figure}
  \epsfig{file=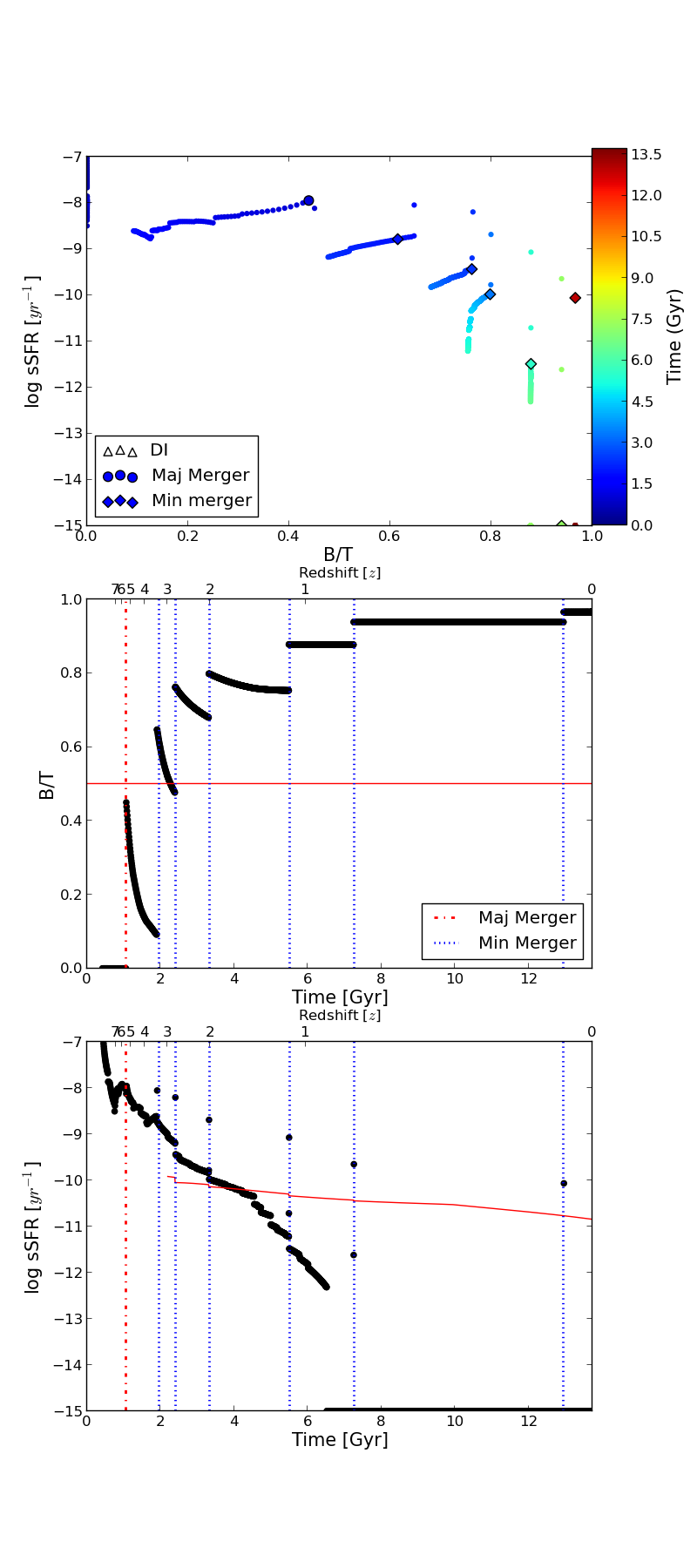, width=0.5\textwidth}
  \caption{Same as previous figure, but for a galaxy which is
    classified as a quiescent spheroid-dominated galaxy at $z=0$
    with a mass of $\sim10^{11.8}M_{\odot}$.  Although this
    galaxy has suffered one major merger at very high redshift
    ($z=5.5$), the build-up of its dominant spheroid occurs through a
    sequence of multiple minor mergers. This is quite typical. The
    build-up of the spheroid is accompanied by growth of the SMBH,
    leading to strong radio-mode feedback that shuts down cooling and,
    eventually, quenching of star formation. } {\label{qetime}}
\end{figure}

Figure \ref{qetime} is a somewhat striking example of a galaxy being
pummeled repeatedly by mergers until it is almost entirely bulge
dominated, after which it finds itself unable to form more stars
because of the black hole it has grown over the course of its
traumatic history; the black hole is now keeping any remaining gas too
hot for star formation through radio-mode feedback. This can also be
effected by one (or more) big major merger(s). In both cases, it is
likely that the system will make an appearance as an SFS for a time
before quickly evolving into the QS quadrant. Once a galaxy falls into
this quadrant, it tends to stay there, except for very rarely when it
collides with a gas-rich galaxy, at which point it might briefly
return to the SFS quadrant before quickly using up all of its new gas
and falling back down again. This can be seen in the very short spikes
of star formation accompanying the last two mergers in the bottom
panel. This galaxy is far more massive than the SFD above, with a stellar mass of $\sim10^{11.8}M_{\odot}$, which is due to all of the merger events it has experienced and is consistent with many of the most massive galaxies in the universe being QSs. We note that the QS population in the models is much more
concentrated toward lower S{\'e}rsic index than that from the
observations, as can be seen in Figure \ref{quadsplits}. Our model is
still not producing enough very bulge-dominated galaxies. As the most
massive bulges are believed to be the result of mergers, this may
indicate that we are still underestimating the role of merging.

\begin{figure}
  \epsfig{file=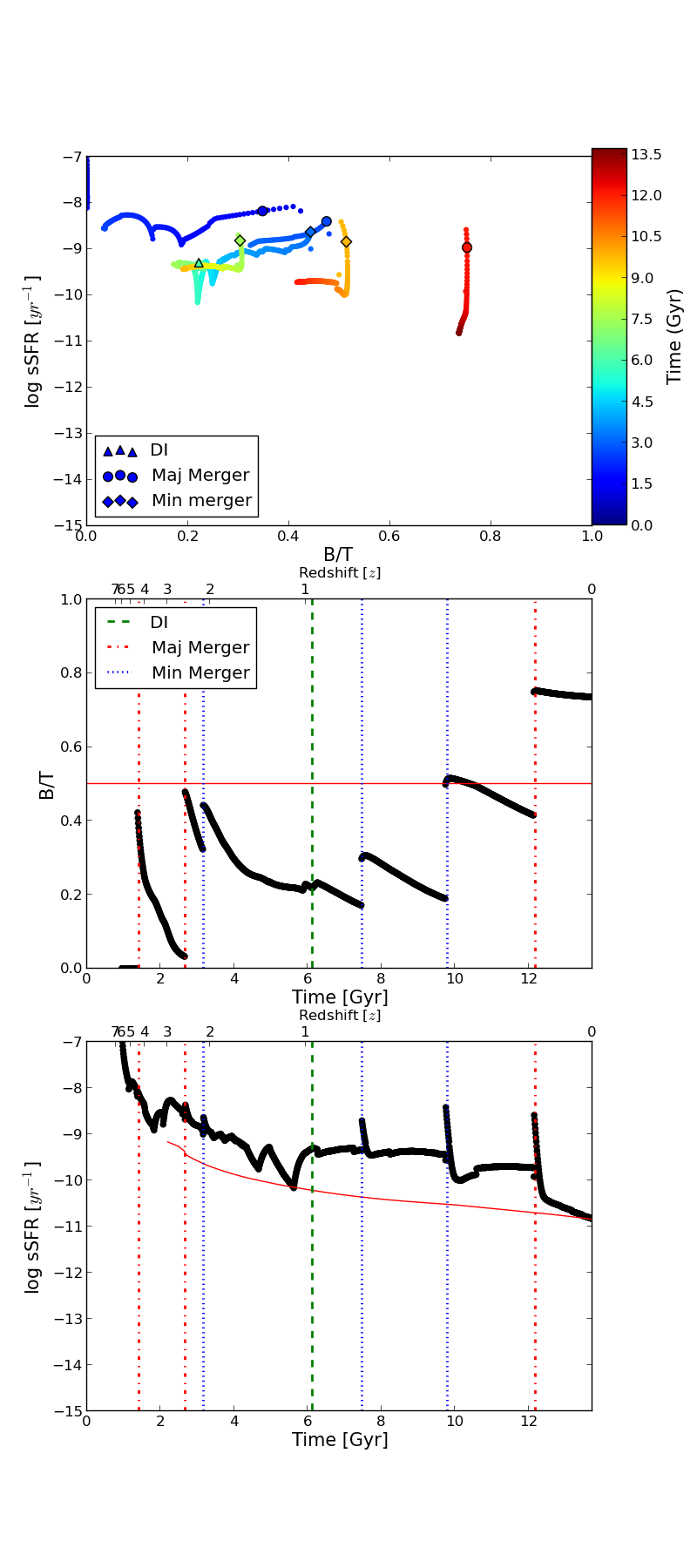, width=0.5\textwidth}
\caption{Same as previous figure, but for a galaxy which is classified
  as a star forming spheroid-dominated galaxy at $z=0$ with a stellar mass of $\sim10^{11}M_{\odot}$. This galaxy
  has had an extremely active history, with multiple major mergers,
  several minor mergers, and a disk instability. A recent major merger
  triggered a strong burst of star formation. This galaxy would
  probably appear morphologically disturbed. } {\label{sfetime}}
\end{figure}

Figure \ref{sfetime} shows the evolution of a galaxy that ends up as
an SFS. Its final mass is $\sim10^{11}M_{\odot}$, which is more massive than the SFD considered above because of its more active merger history. We see here what a major merger can do in terms of bulge
growth: the first major merger this galaxy experiences gives it a
substantial bulge component. After each bulge growth episode, the
galaxy begins to regrow a disk, causing $B/T$ to decrease steadily. This
is the case for many galaxies in the SAM, as long as they have the gas
to form new stars or continue to accrete new gas from the IGM. Changes
in sSFR between merger events for this system are largely due to the
interplay between ``normal'' star formation and new gas accretion. If
this galaxy had not undergone a major merger recently, it would not
have been considered an SFS, as the steady regrowth of its disk would
have caused it to be classified as disk-dominated instead.

The SFS quadrant is more of a way station than a destination in galaxy
evolution, with galaxies either quickly evolving back towards the SFD
quadrant by regrowing their disks, or evolving downwards into the QS
quadrant as they quench (this system is very close to being classified
as a QS). This can be seen in Figure \ref{quadsplits}, where the SFS
quadrant appears more to be made up of the tails of populations in the
SFD and QS quadrants than to be a distinct population. This is
something upon which the models and observations appear to agree, and
is an explanation for why the evolution of the overall
spheroid-dominated fraction is less steep than for the quiescent
fraction: there is high turnover in the SFS quadrant so the fraction
in that state is relatively constant, meaning that the buildup of the
spheroid-dominated population relies mainly on the steady buildup of
QSs. Meanwhile the quiescent fraction is built up by the steady growth
of both the QS and QD populations. This interpretation seems to be
corroborated by the CANDELS-based study of Rizer et al. (in prep.), in
which visual morphological classifications are used. They find that
while their QS population builds up steadily, their SFS population
remains relatively constant, suggesting that bulge growth is leading
to star formation quenching in many cases.

\begin{figure}
  \epsfig{file=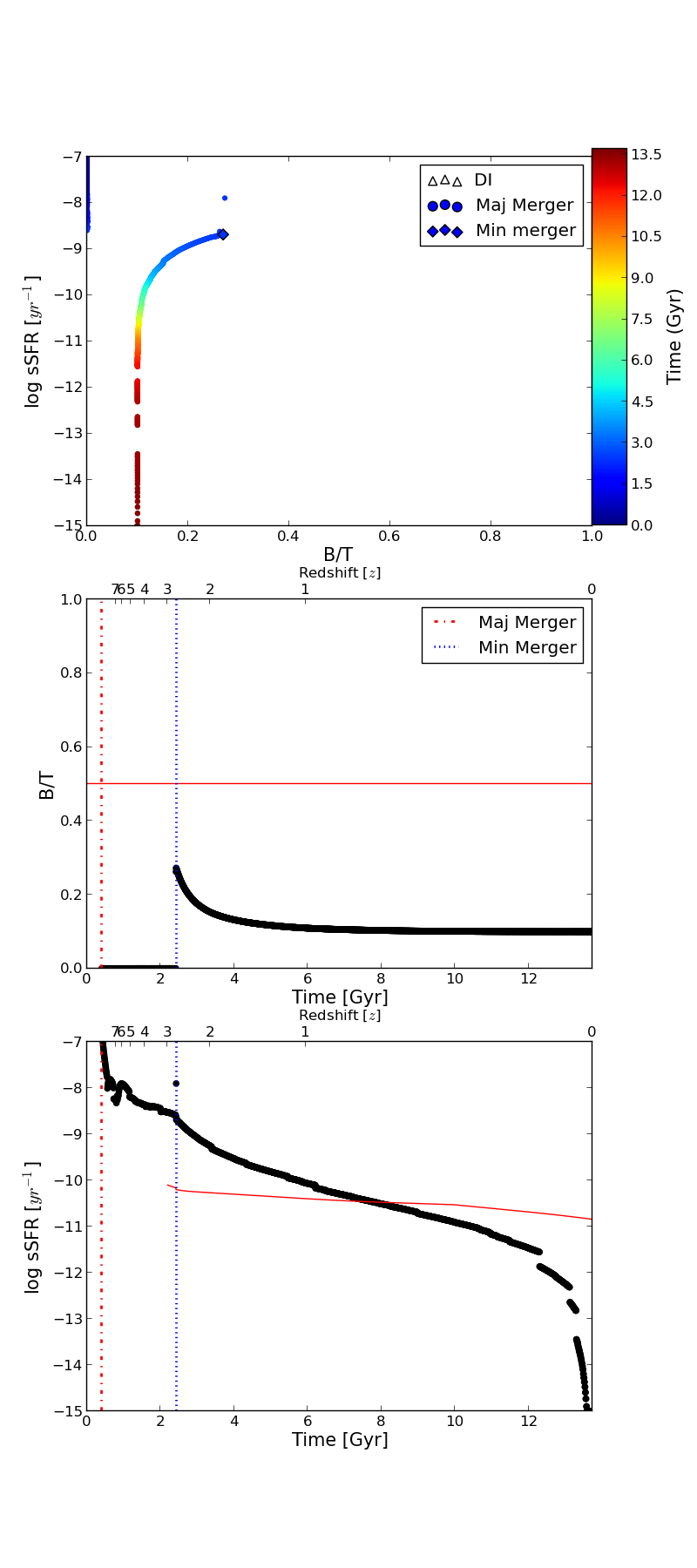, width=0.5\textwidth}
  \caption{Same as previous figure, but for a galaxy which is classified as a quiescent disk-dominated galaxy at $z=0$ with a stellar mass of $\sim10^{11.7}M_{\odot}$.}
  {\label{qdtime}}
\end{figure}

Figure \ref{qdtime} represents one possible path to becoming a
quiescent disk-dominated galaxy. Some quiescent disks in our model are
galaxies with quiet histories such as the one seen in Figure
\ref{sfdtime} above, but which have run out of gas (perhaps because
they are more massive and have a harder time accreting new
material) or can't form stars with the gas they do have (see below for a brief discussion). However, many of our disk-dominated quiescent galaxies
really aren't very disk-dominated but in fact are systems which have
had a somewhat more eventful history similar to that of a QS. These
end up as QDs by one of two ways. In the first case, the events which
lead to star formation quenching and bulge formation do not form
enough of a bulge for the system to be considered
spheroid-dominated. However, the galaxy still falls into the QD
quadrant in the same way we expect SFSs to migrate sometimes to the QS quadrant
after a traumatic event. In the second case, the galaxy \textit{does}
become an SFS, but retains gas to regrow a disk sufficient to be
considered disk-dominated before falling into quiescence. 

We see a mixture of these two fates in Figure \ref{qdtime}; while its
bulge growth episode was not enough to make it spheroid-dominated, the
event which caused the bulge growth was enough to cut this galaxy off
from new gas, causing it to become quiescent. We also see substantial
disk growth in its decreasing $B/T$. This combination of both
scenarios causes this galaxy to be somewhat rare: a QD which is both
very disk-dominated and very quiescent. This galaxy is very
  massive at $z=0$ with a stellar mass of
  $\sim10^{11.7}M_{\odot}$. QDs are often quite massive due either to
  their merger histories or prolonged star formation which has led the
  galaxy to be unable to accrete new gas due to virial shock heating.
In general, the average $B/T$ for QDs is only slightly higher than for
SFDs. However, as QDs with lower star formation rates are considered,
the average $B/T$ becomes larger, as these systems are the result of
the QS-like paths described above. The very disk-dominated QDs are
likely to have relatively higher sSFRs. This does bring up a larger
point about dividing lines in general: some galaxies which fall below
our dividing line are still forming stars, albeit at a slower rate
than the majority of galaxies of their mass. They are ``quenched'' in
the sense that their star formation rate is lower than expected, but
they are not truly ``quiescent'' as is the case for some of our
galaxies, which have had their star formation completely turned
off. There is also the possibility, mentioned above, that some of our
quiescent galaxies might begin to form stars again due to a wet merger
or new gas accretion. In the case of a wet merger, this is likely to
be short-lived, but in the case of new gas accretion, it can lead to a
whole new life for a galaxy. Having clarified that, however, we
believe this is happening in both the models and the observations, so
it should not bias our results.

It appears that the QDs are much like the SFSs in that the population
is a combination of tails of the SFD and QS populations. Curiously,
the gas fractions of QDs in the SAMs are \emph{not} systematically
smaller than those of SFDs of the same mass. It appears that the
quenching leading to very disk-dominated QDs has two possible origins,
as mentioned above: a low gas accretion rate, or an extended low
surface-density gas disk which is inefficient at making stars. In the
models considered here, only gas that is above a critical surface
density is allowed to participate in star formation; galaxies with
larger than average angular momentum form more extended disks, which
have a larger fraction of their gas sitting below this critical
surface density. There is observational evidence in the local universe
for these gas rich but relatively quiescent disks
\citep{Lemonias:2014,Schiminovich:2010}. Thus while the SFSs may be
understood as a transient population, a step in the path from SFD to
QS, it seems that QDs may be a static population.

In addition, our model QDs on average have smaller stellar and black
hole masses than our QSs. They are also more likely to be satellites
than QSs. In our lowest redshift bin, about 75\% of model QSs are
central galaxies, while only 50\% of QDs are. The \textit{very}
disk-dominated QDs ($B/T<0.1$) don't exist in our model before
$z\sim2$. When compared with the rest of the QDs, they have even
smaller black hole masses, as they likely have not undergone any
events which would have triggered AGN feeding. They are also even more
likely to be satellites; in our lowest redshift bin, only 25\% of very
disk-dominated QDs are central galaxies. We note that the likelihood
of being a satellite galaxy is not the only reason that QDs tend to
have smaller stellar and black hole masses; this trend is observed
even when only central galaxies are considered.

What these analyses and evolutionary tracks show is that the
transformative processes which affect galaxies take them all over the
map (and the sSFR-$n$ plane). It is likely too simplistic to tell a
simple story about two star forming disk galaxies colliding or one of
them buckling under its own weight and triggering feedback which
produces a nice, dead elliptical galaxy. These processes (mergers,
disk instabilities, accretion of new gas) likely work together,
sometimes in tandem and sometimes at cross purposes. It appears that a
complex history with multiple transformative events is the norm rather
than the exception and galaxy histories don't necessarily look like
the arrows in Figure \ref{quad}.

\subsection{\emph{How Might a Different Implementation of Disk Instability Change Our Results?}}

As mentioned above, our treatment of disk instabilities is based on
rather dated simulations of isolated disk galaxies which are not in a
cosmological context, and may not capture all of the relevant
physics. Some of the questions associated with disk instabilities
include: 1) What is the most relevant criterion for determining the
onset of a disk instability? 2) What happens to the gas and stars in
the disk when it becomes unstable? 3) How efficiently do disk
instabilities feed a nuclear black hole?

As an aside, we have so far elided over a possibly important
distinction. There are two kinds of physical mechanisms that are
commonly referred to as ``disk instabilities'' in the literature,
although in one case this is something of a misnomer, as we explain
below.  In ``violent disk instabilities'' \citep[VDI;][]{Dekel2009},
the disk becomes globally unstable, leading to the formation of clumps
of stars and gas, which may migrate to the center of the galaxy,
building the spheroid.  As the giant clumps orbit within the disk,
even if they disrupt before reaching the center, they may drive
inflows of gas into the galaxy nucleus, via the same sort of physics
as merger-induced nuclear inflows, again leading to growth of the
spheroid through in situ star formation, and feeding of the black hole
\citep{Bournaud2011}. The second kind of ``disk instability''
involves the secular transfer of angular momentum outwards, and mass
inwards, again leading to the building of a central compact and
dynamically hot structure, and is accompanied by the formation of a
bar \citep{Kormendy:2004}. The term ``instability'' is a misnomer
here, as the disk essentially remains in dynamical equilibrium. The
relatively crude morphological statistics used in this analysis are
not able to distinguish between edge-on bars and bulges, so some of
the `spheroids' we count in the observations may actually be bars. In
addition, the nuclear structures that are formed via this secular
process do not have the same properties as ``classical'' bulges, and
are sometimes called ``pseudobulges''. See \citet{Kormendy:2004} for a
detailed discussion of the differences between classical bulges and
pseudobulges; the most germane for our purposes here is that
pseudobulges do not obey the same scaling relationships with the SMBH
mass as classical bulges \citep{Kormendy:2013}, suggesting that black
hole feeding and/or feedback may operate differently.

One major limitation of our approach is that we based the disk
instability criterion on the properties of the stellar disk only, and
only stars are moved from the disk to the spheroid when the disk is
deemed unstable. This is not what is seen in modern cosmological
hydrodynamic simulations, in which VDI are ubiquitous at high redshift
\citep{Ceverino:2010,Mandelker:2014}, and as mentioned above are
associated with strong nuclear inflows of gas as well as stars. In
addition, VDI can lead to significant quenching even in the absence of
associated AGN feedback \citep{Gammie2001,Dekel2014,Forbes2014}, which
is not accounted for in our current models. Perhaps this could lead to
a higher fraction of quiescent galaxies at high redshift, in accord
with observations. Additional complications are the possible
stablizing effects of a pre-existing bulge or central mass
concentration, and possible triggering of disk instabilities by minor
mergers. 

P14 attempted to model a scenario closer to the modern VDI picture
with their \textbf{``Stars+Gas DI''} model. They found that the
results were very similar to the \textbf{``Stars DI''} model which is
why we consider only that model here (this implementation of DI is
similar to the one that is most commonly applied in other SAMs in the
literature). In addition, the P14 \textbf{``Stars+Gas DI''} model is
still very arbitrary and simplified. In future work, we plan to carry
out a thorough analysis of state-of-the-art hydrodynamic simulations
to develop a more detailed and physical treatment of disk
instabilities in SAMs.

\subsection{\emph{Other Possible Model Improvements}}
Aside from our treatment of disk instabilities, there are several ways
in which we could improve our model in order to better capture what we
believe is occuring in real galaxies. Our implementation of satellite
stripping does not account for any change in morphology of satellites
and many which are not destroyed are stripped of their gas and remain
disk dominated. Only central galaxies are supplied with new gas in our
accretion and cooling model, so this may artificially increase our
fraction of quiescent disk-dominated galaxies. While this effect
dominates at low stellar masses and should not be as important in the
mass range we consider in this work, $M_{*}>10^{10}M_{\odot}$, it is
worth noting (and is the reason we neglected satellites when
determining the typical model main sequence star formation rate
above). We do point out that the same analysis done for the model
excluding satellite galaxies leads to very similar results; the
largest difference is in the low redshift QD fraction, which in the
lowest bin decreases from $\sim25$\% to $\sim17-18$\%.

While our merger prescription is based on numerical hydrodynamic
simulations of binary mergers \citep{Hopkins2009a}, we treat all
mergers as discrete events, while in the real universe there may be a
complex interplay between merger events if the galaxy has not had time
to relax between them \citep{Moster2014}. This could have an effect on
both the star formation rates and the morphologies of our post-merger
remnants, especially of those that have had somewhat active merger
histories.

In addition, in our current models galaxies are primarily kept
quiescent by the ``maintenance mode'' type radio mode feedback, which
becomes important only at relatively late times. Winds driven by
radiatively efficient accretion (``bright mode'' feedback) are assumed
to be able to remove cold gas from the ISM but have no effect on the
hot gas surrounding galaxies. However, recent cosmological simulations
have shown that momentum-driven winds associated with bright mode
accretion actually modify the hot gas profile and significantly retard
cooling over long timescales \citep{Choi:2014a,Choi:2014b}. Including
this physics might also help us to produce more quenched galaxies at
high redshift, and could also suppress the re-formation of disks via
cooling.

Finally, there is the issue of putting the models and observations on
the same footing when it comes to quantitative comparison of
morphologies; we would like to improve the way we assign morphology to
our model galaxies, namely by using the masses (or luminosities) and
sizes of our disk and bulge components to generate mock images which
can then be processed like real observations (including noise,
point-spread function, etc) and assigned morphological
classifications. This way each galaxy would have its profile measured
separately and other effects such as inclination angle could be taken
into account. Comparing the observed and predicted fractions of
morphologically disturbed galaxies would also be interesting, but
obviously requires us to be able to quantify this in the SAM in a way
that can be compared with observations. This may be possible using our
merger statistics in combination with a library of numerical
simulations, as in \citet{Lotz2011}; see also recent work by
\citet{Snyder2014}.

\section{Summary and Conclusions}
We have studied the coevolution of star formation rate and morphology
from $z\sim3$ to the present by examining the buildup of galaxies in
the four quadrants of the sSFR versus S{\'e}rsic index ($n$) plane. We
have compared galaxies with stellar mass >$10^{10}M_{\odot}$ from the
``Santa Cruz'' semi-analytic model outlined in S08, S12 and P14 with
galaxies observed as part of the GAMA and CANDELS surveys. Our
conclusions are as follows:

\begin{itemize}
\item Our models qualitatively reproduce the increasing fraction of
  quiescent galaxies since $z\sim 2$ seen in observations, and produce
  excellent quantitative agreement with observations at $z\lesssim
  1.2$. At higher redshift, the models underproduce the fraction of
  quiescent galaxies relative to observations.

\item Our model in which spheroids are built solely through mergers
  (noDI) predicts an evolution in the fraction of spheroid dominated
  galaxies that is much too mild compared with observations. This
  model also underproduces spheroid-dominated galaxies at $z \lesssim
  2$ compared with observations. Adding a channel for bulge growth via
  disk instabilities (DI model) leads to much better agreement with
  the observed evolution of the spheroid-dominated fraction, although
  still produces slightly flatter evolution than observed.

\item The quiescent fraction is largely unaffected by the inclusion
    of disk instabilities, with changes <10\%. We note, however, that
    our current disk instability model may be underestimating the
    change in quiescent fraction, especially at high redshift.
  
\item Our models further qualitatively reproduce the observed
  evolutionary behavior of four classes of galaxies defined by both
  star formation activity and morphology: star forming disk-dominated,
  quiescent spheroid-dominated, star forming spheroid-dominated and
  quiescent disk-dominated. In both the observations and in our
  models, the fraction of star forming disks decreases over time while
  the fraction of quiescent spheroids increases. In the observations,
  the fractions of both star forming spheroids and quiescent disks
  remain nearly constant from $z\sim 3$--0. Models predict a stronger
  decrease in star forming spheroids and a stronger increase in
  quiescent disks with redshift than is seen in the observations, but
  the predicted fractions are not off from the observed fractions by more than $\sim0.2$ and in most cases are off by less.

\item In our models, star forming-disk dominated galaxies are
    galaxies which have had very quiet histories. They have avoided
    major mergers and if they have experienced any merger or disk
    instability activity, they have recovered by accreting new gas and
    regrowing a disk.

\item Again in our models, quiescent spheroid-dominated galaxies are
  likely to have either undergone at least one extreme major merger or
  many smaller mergers or disk instabilities. In either case they have
  built up a substantial bulge component and AGN feedback has made it
  impossible for them to accrete significant amounts of new gas,
  eventually leading to cessation of star formation.

\item Star forming spheroid-dominated galaxies seem to be a
  short-lived population. Truly spheroid-dominated star forming
  galaxies are indicators of a recent trauma, as they are still
  experiencing a post-trauma starburst. At this point, they can regrow
  a disk with their remaining gas reservoir or through the accretion
  of new gas. In the absence of new gas, they can deplete their gas
  reservoirs and become quiescent.

\item Quiescent disk-dominated galaxies are a combination of two
   populations: disk-dominated galaxies which have stopped accreting
   gas (in some cases due to environmental effects) and galaxies with
   extended low-surface density gas disks, which are inefficient at
   forming stars.

\end{itemize}

Despite the room for possible improvements to our model described
above, the qualitative similarity between the buildup of our model
populations with those of observed galaxies gives us confidence that
we are beginning to capture the complicated interplay of several
processes which lead to the diversity of galaxies and their evolution
over time.

\section*{Acknowledgments}
We thank the anonymous referee for a very positive and helpful report. RB was supported in part by HST Theory grant HST-AR-13270-A. rss thanks the Downsbrough family for their generous support. We acknowledge the contributions of hundreds of individuals to the planning and support of the CANDELS observations, and to the development and installation of new instruments on HST, without which this work would not have been possible. Support  for HST Programs  GO-12060  and  GO-12099  was provided by NASA through grants from the Space Telescope Science  Institute,  which  is operated  by  the  Association  of Universities  for  Research  in Astronomy,  Inc.,  under  NASA contract NAS5-26555.

\bibliographystyle{mn2e}
\addcontentsline{toc}{section}{\refname}\bibliography{paper}

\section*{Appendix A: Conversion from B/T to S{\'e}rsic Index}

As part of this work, we compare the morphologies of model galaxies to
observed galaxies. To do this, we convert bulge-to-total stellar mass
ratios ($B/T$), which our model naturally outputs, to S{\'e}rsic indices
from single component fits, which are directly comparable to
observations. To do this, we use a lookup table generated from
synthetic galaxies which are made up of an exponential ($n=1$) disk
and a bulge with $n$=4. A S{\'e}rsic index and effective radius are
derived by fitting to the two-component profile for a wide range of
$B/T$ and $r_{\rm{bulge}}$/$r_{\rm{disk}}$ (which we hereafter refer to
as ``rbd''). The lookup table takes in $B/T$ and rbd and outputs a
S{\'e}rsic index and effective radius for the composite system. Since
the values are discrete, we have interpolated between the table values
to generate our S{\'e}rsic indices. Appendix A of \citet{Lang2014}
illustrates that the relationship between $B/T$ and $n$ derived from
these synthetic galaxies matches well with the relationship derived
from CANDELS galaxies with 2-component bulge+disk fits. Here we
present the mapping between $B/T$ and S{\'e}rsic index in order to
illustrate the relationship between the two. Then, in the next
section, we will show that the results of the analysis presented in
the main text are largely unchanged when done in terms of $B/T$ rather
than S{\'e}rsic index. We also refer the reader to Figure A6 in
\citet{Lang2014}, where this analysis is done for the observed
galaxies with bulge-disk decompositions. They carry out their analysis
in terms of both bulge-to-total stellar mass ratio and H-band light
ratio. We have used the lookup table generated in terms of the mass
ratio, but we would expect results in terms of light ratio to be
qualitatively similar (see Appendix B of \citealt{Lang2014}).

\begin{figure*}
  \epsfig{file=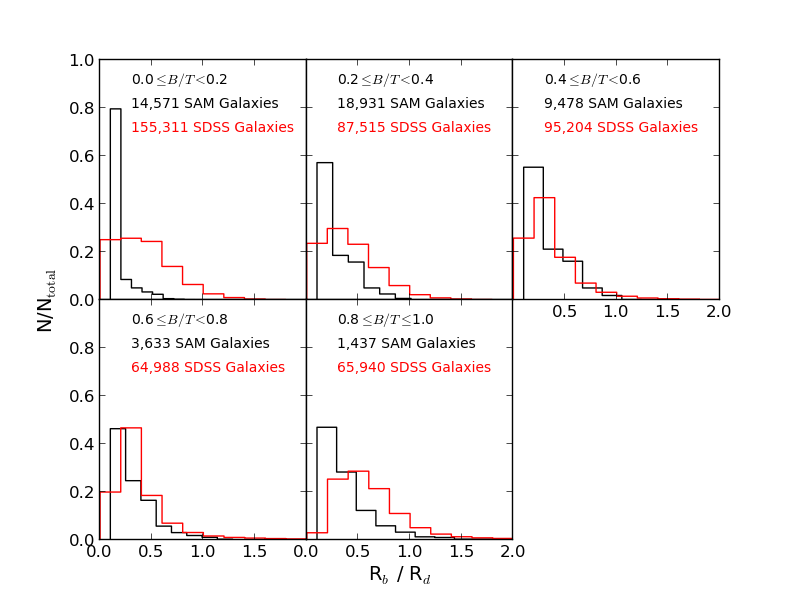, width=0.75\textwidth}
\caption{Distribution of rbd values for SAM galaxies with
  $0.06<z<0.12$ and galaxies from SDSS with bulge-disk decompositions
  from \citet{Simard2011} in bins of $B/T$.}  {\label{rbddist}}
\end{figure*}

To test the SAM predictions for the bulge and disk sizes, we use the
$r$-band bulge+disk decompositions of SDSS galaxies performed by
\citet{Simard2011}. We trimmed the original catalog of 1,123,718
galaxies down to 618,186 galaxies by applying the following selection
cuts: $0.005<z<0.12$, $0.0\leq B/T\leq 1.0$, $0.5\leq n_{\rm pure}\leq
8.0$, $M_r > -99$, $M_{\rm r,err} > -99$, $M_{\rm r,pure} > -99$,
$M_{\rm r,pure,err} > -99$, $r_{\rm bulge,eff} > 0$, $r_{\rm disk,eff} > 0$, and
$r_{\rm pure,eff} > 0$, where the subscript ``pure" refers to
single-component (pure) S{\'e}rsic fits (which were also computed for the
galaxies). The bulge+disk decompositions were fit simultaneously in
the $r$- and $g$-band in order to minimize errors, and the assumed
model was a de Vaucouleurs bulge ($n_{\rm bulge}=4$) with a pure
exponential disk. The fits were done using the Galaxy Image 2D (GIM2D)
program; see \citet{Simard2002} and \citet{Simard2011} for further
details about the fitting procedure and outputs. In general, the model
predictions and observational results are similar, except in the
lowest $B/T$ bin, where our model predicts more compact bulges relative
to the disk sizes than is seen in the observations. This excess is
seen to a lesser degree in the other $B/T$ bins as well. However, as the
discrepancy is the largest for disk-dominated galaxies (where the
radial size of the bulge component will have little impact on our
results), we conclude that our model should produce reasonable
predictions for the composite S{\'e}rsic indices in most cases. In the
future, it will be interesting to compare the SAM results with the
sizes and $B/T$ ratios obtained from multi-component bulge-disk
decompositions, which are starting to become available.

\begin{figure*}
  \epsfig{file=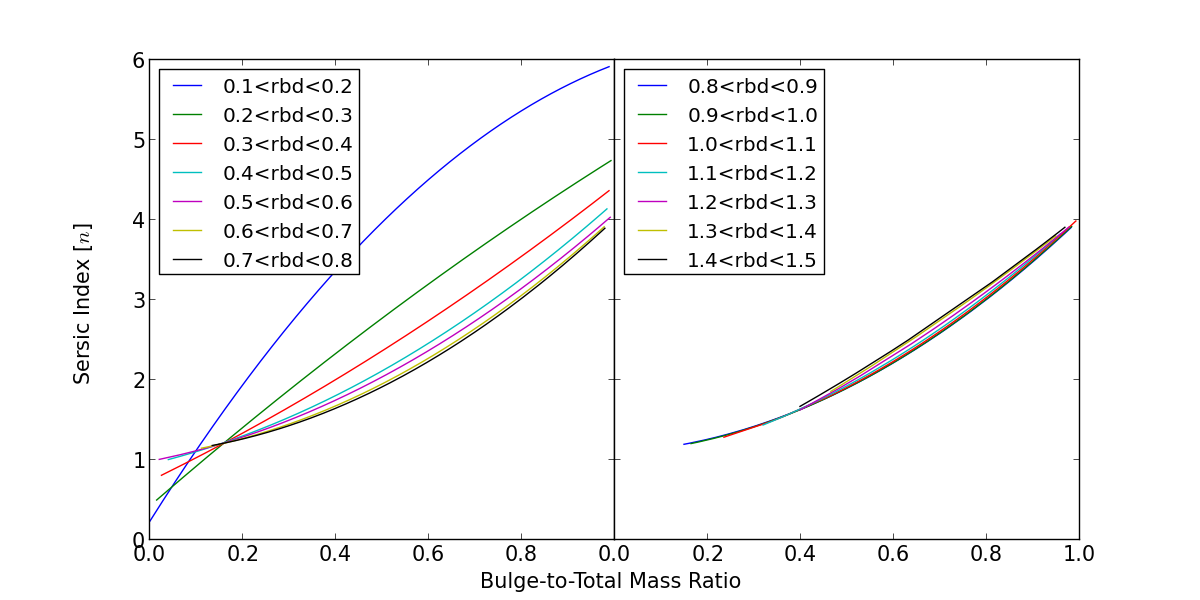, width=0.75\textwidth}
  \caption{Fit to the relationship between $B/T$ and S{\'e}rsic index in
    bins of bulge radius/disk radius. For compact bulges, the
    S{\'e}rsic index is a function of both $B/T$ and rbd. As rbd
    increases, the relationship becomes degenerate.}  {\label{btmap}}
\end{figure*}

Figure \ref{btmap} shows the best fit curves to the relationship
between $B/T$ and $n$ in bins of rbd. If rbd$<0.4$, for a given $B/T$ a
larger rbd will lead to a lower S{\'e}rsic index (unless the galaxy
also has very low $B/T$, in which case $n$ is mostly concentrated
between $\sim0.5-1.0$ anyway). Above an rbd of $\sim0.4$, there is a
nearly one-to-one mapping between $B/T$ and S{\'e}rsic index. However,
as we saw in Figure \ref{rbddist}, in the SAMs and in nearby galaxies,
most galaxies have rbd$<1$. We see here that a split in $B/T$ instead
of $n$ will lead to the selection of slightly different sets of
galaxies because the bulge radius to disk radius ratio causes a spread
in S{\'e}rsic index for a given bulge-to-total mass ratio and vice
versa.

\section*{Appendix B: Results Using B/T}

\begin{figure}
  \epsfig{file=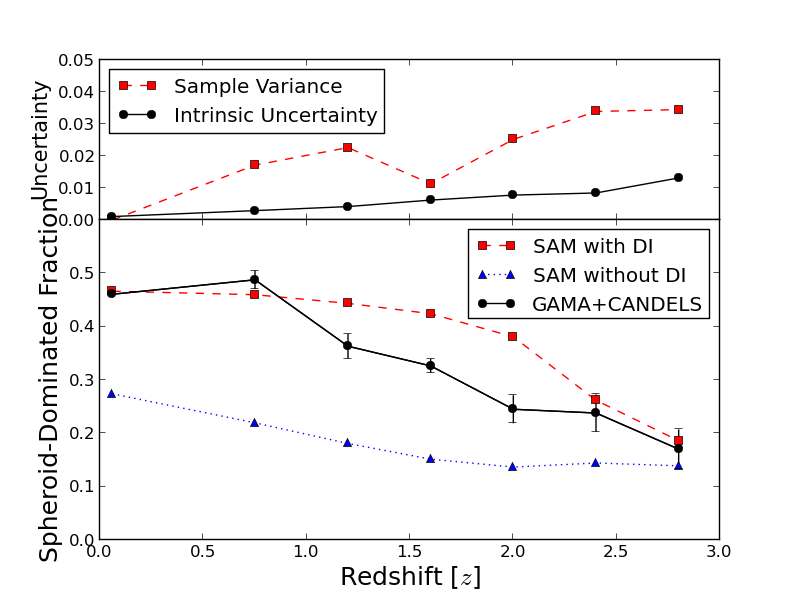,
    width=0.5\textwidth}
\caption{The evolution of the spheroid-dominated fraction of galaxies
  with redshift, now with spheroid domination defined as
  $B/T>0.5$. The observed galaxies are still split by S{\'e}rsic index
  at $n=2.5$. Error bars are the $1-\sigma$ uncertainties due to
  sample variance in the models and uncertainty in observed galaxy
  properties added in quadrature. The separate contributions are
  plotted in the top panel.}  {\label{efracbt}}
\end{figure}

In this section we present the main (morphology-dependent) results of
the above analysis again, this time using $B/T$ as our morphological
parameter. These results may be able to be compared with future
observational analyses, if bulge-disk decompositions are carried out,
and may be more easily compared with predictions from other
theoretical models.  Our dividing line between star forming and
quiescent remains the same, but our condition for spheroid domination
is now $B/T>0.5$. Figure \ref{efracbt} shows the evolution of the
spheroid-dominated fraction of galaxies as in Figure \ref{efrac}. The
observed spheroid-dominated fraction is still derived using a
S{\'e}rsic index $n=2.5$ to make the cut. We can see that the
evolution of model galaxies is similar to the evolution in Figure
\ref{efrac}. However, splitting by $B/T$ we predict more
spheroid-dominated galaxies for $z<2$ and fewer for $z\sim3$. The
predicted fraction does not vary by more than 0.1 at any
redshift. This variation suggests that there are more galaxies at high
redshift with smaller values of rbd than at low redshift.

\begin{figure*}
  \epsfig{file=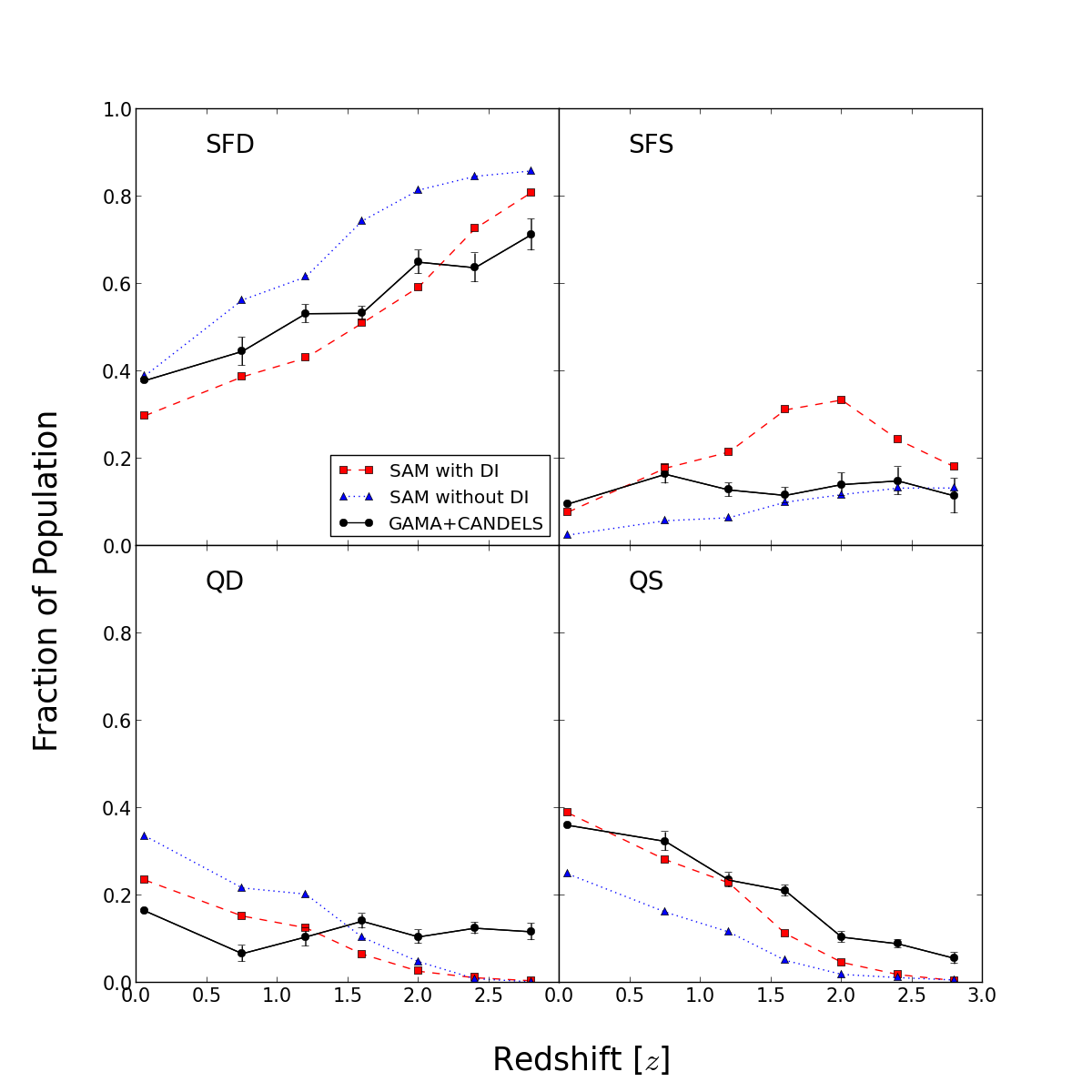, width=1.0\textwidth}
  \caption{Same as Figure \ref{quadevol}, but now with model galaxies
    split by $B/T=0.5$. Observed galaxies are still split by
    S{\'e}rsic index at $n=2.5$. Top left: Star forming disk-dominated
    galaxies. Top right: Star forming spheroid-dominated
    galaxies. Bottom left: Quiescent disk-dominated galaxies. Bottom
    right: Quiescent spheroid-dominated galaxies.}
          {\label{quadevolbt}}
\end{figure*}

\begin{figure*}
  \epsfig{file=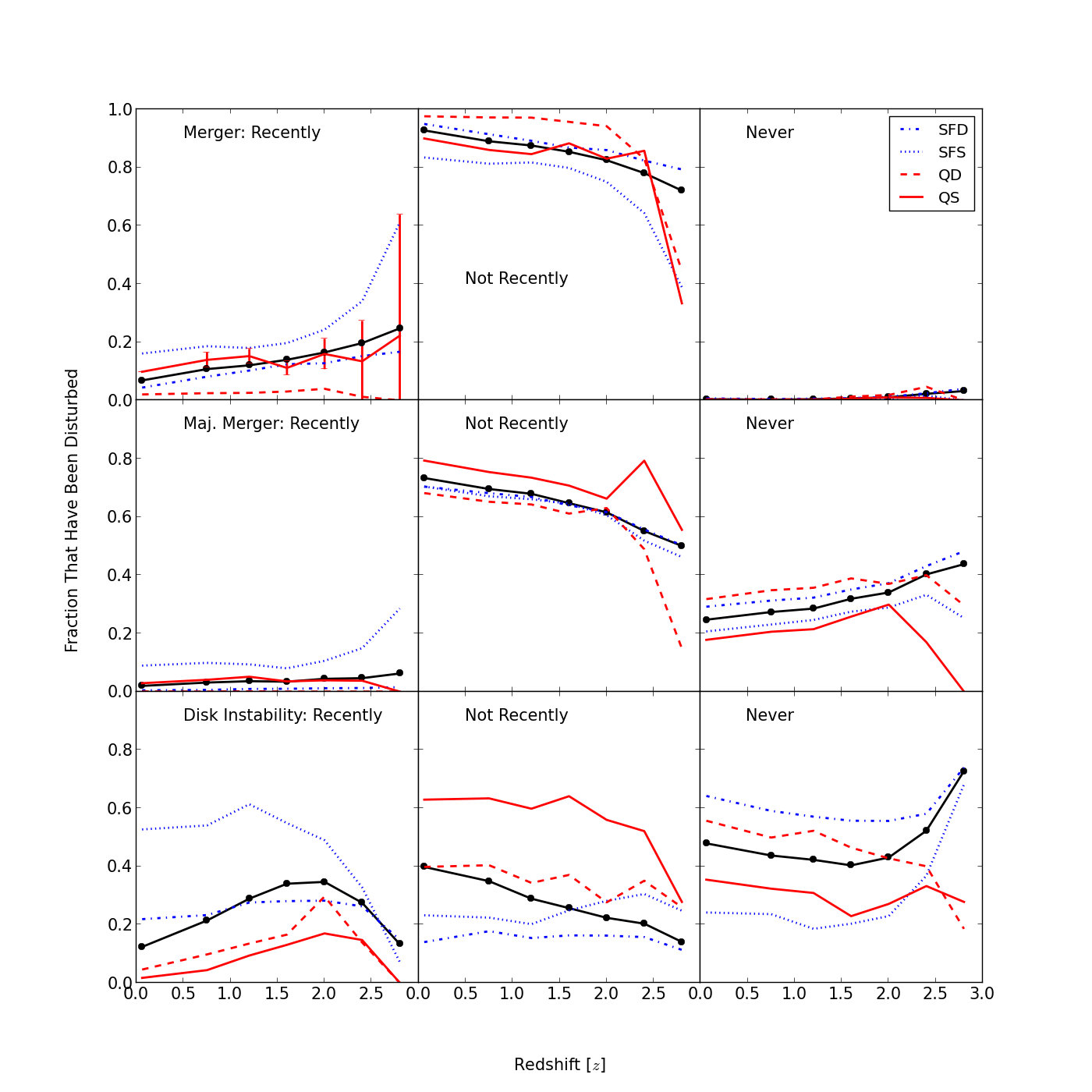, width=1.0\textwidth}
  \caption{Same as Figure \ref{tmfracevol}, but with a morphology cut
    at $B/T=0.5$. Top row: Left panel: Fraction of galaxies in each
    quadrant which have undergone a recent (<$3t_{\rm{dyn}}$) merger
    with mass ratio >1:10. Middle panel: Fraction which have undergone
    a merger on a timescale >$3t_{\rm{dyn}}$. Right panel: Fraction
    which have never undergone a merger. Fractions are now determined
    using the $B/T$ cut. Middle row: Same as top row, but for major
    mergers with mass ratio >1:3. Bottom row: Same as top and middle rows but for disk
    instabilities.}  {\label{tmfracevolbt}}
\end{figure*}

The evolution of the quadrant fractions in Figure \ref{quadevolbt} is
extremely similar to that in Figure \ref{quadevol}. The fraction
predicted in each of the spheroid-dominated quadrants is slightly
larger than in the S{\'e}rsic index case. Figure \ref{tmfracevolbt} is
like Figure \ref{tmfracevol}. The only significant difference when
splitting by $B/T$ instead of $n$ is in the disk instability plot. More
SFSs and QSs have had recent disk instabilities as defined by $B/T$ than
by $n$. This is because the change in definition is not likely to
affect galaxies that are very clearly disk-dominated or
spheroid-dominated. The galaxies that have intermediate S{\'e}rsic
indices or $B/T \sim0.5$ are the ones that are traded back and forth
depending on definition (because of their rbd values), and a large
fraction of these are created by the disk instability. This is also
why the results for the DI model seem to be affected more strongly by
morphological definition than the noDI model.

\end{document}